\documentclass[12pt, draftclsnofoot, onecolumn]{IEEEtran}

\usepackage{cite}
\usepackage{datetime}
\usepackage{graphicx}
\usepackage{subcaption}
\usepackage{epstopdf}
\usepackage{amsmath}
\usepackage{amssymb}
\usepackage{amsfonts}
\usepackage{amsthm}
\usepackage{algorithmic}
\usepackage{array}
\usepackage{stfloats}
\usepackage{multirow}
\usepackage{color}
\usepackage{placeins}
\usepackage{gensymb}
\usepackage{array}
\usepackage{dsfont}
\usepackage[font=small]{caption}
\usepackage[hidelinks]{hyperref}

\newcommand\abs[1]{\left|#1\right|}

\begin{document}

\title{Downlink and Uplink Cell Association with Traditional Macrocells and Millimeter Wave Small Cells}

\author{\IEEEauthorblockN{Hisham Elshaer, Mandar N. Kulkarni, Federico Boccardi, Jeffrey G. Andrews, and Mischa Dohler} \thanks{This work has been supported by Vodafone Group R\&D and CROSSFIRE MITN Marie Curie project. H. Elshaer (\href{mailto:hisham.elshaer@vodafone.com}{\texttt{hisham.elshaer@vodafone.com}}) is with Vodafone Group R\&D and King's College London, UK. M. N. Kulkarni (\href{mailto:(mandar.kulkarni@utexas.edu}{\texttt{mandar.kulkarni@utexas.edu}}) and J. G. Andrews (\href{mailto:jandrews@ece.utexas.edu}{\texttt{jandrews@ece.utexas.edu}}) are with the Wireless Networking and Communications Group (WNCG), The University of Texas at Austin, USA. F. Boccardi (\href{mailto:federico.boccardi@ofcom.org.uk}{\texttt{federico.boccardi@ofcom.org.uk}}) is with OFCOM, UK. M. Dohler (\href{mailto:mischa.dohler@kcl.ac.uk}{\texttt{mischa.dohler@kcl.ac.uk}}) is with the Centre for Telecommunications Research (CTR), King's College London, UK. }}

\IEEEtitleabstractindextext{%
\begin{abstract}
Millimeter wave (mmWave) links will offer high capacity but are poor at penetrating into or diffracting around solid objects.  Thus, we consider a hybrid cellular network with traditional sub 6 GHz macrocells coexisting with denser mmWave small cells, where a mobile user can connect to either opportunistically.  We develop a general analytical model to characterize and derive the uplink and downlink cell association in view of the $\mathtt{SINR}$  and rate coverage probabilities in such a mixed deployment.  We offer extensive validation of these analytical results (which rely on several simplifying assumptions) with  simulation results.  Using the analytical results, different decoupled uplink and downlink cell association strategies are investigated and their superiority is shown compared to the traditional coupled approach. Finally, small cell biasing in mmWave is studied, and we show that unprecedented biasing values are desirable due to the wide bandwidth.
\end{abstract}

}

\maketitle

\IEEEdisplaynontitleabstractindextext

%
\IEEEpeerreviewmaketitle

\section{Introduction}
Two key capacity-increasing techniques for future cellular networks including 5G will be network densification and the use of higher frequency bands, such as millimeter wave (mmWave) \cite{AndBuz14, BocHea14}.  The main challenges to using mmWave frequencies are their high near-field pathloss (due to small effective antenna aperture) and very poor penetration into buildings.  However, it is increasingly believed these challenges can be overcome, at least for outdoor-to-outdoor cellular networks, using high gain steerable antennas in a dense enough network with sufficient scattering \cite{Rap13,PiKha11,RohSeo14, RapGut13,RanRap14,Akd14,LarTho13,Gho14,BaiHea14}.  Further, recent studies have shown that  with such highly directional transmissions and sensitivity to blockage, a positive side effect is that interference is greatly reduced, and so  in many or most cases,  mmWave networks will be noise rather than interference-limited \cite{BaiHea14, SinKul14,Gho14, Ren15}.   Nevertheless, it is unrealistic to expect universal coverage with mmWave, especially indoors, and so  a likely deployment scenario is that mmWave will co-exist with a traditional sub-6GHz cellular network.  The mmWave small cells will be used opportunistically when a connection is possible, with the sub-6GHz base stations providing universal coverage, for both control signaling and for data when a mmWave connection is not available. 
  The goal of this paper is to model and analyze such a hybrid network, considering in particular how user equipments (UEs) should associate with the two types of BSs in the uplink (UL) and downlink (DL).

\subsection{Related Work}
Downlink and uplink associations are typically coupled, i.e. a UE connects to the same BS in the DL and UL.  In the context of a heterogeneous network, downlink-uplink decoupling (DUDe) has been recently shown to significantly improve the network capacity (especially in the UL) by considering different association criteria for the UL and DL \cite{ElsBoc14}.   DUDe has been discussed in \cite{BocHea14, JA2013, AstDah13} as an interesting component for future cellular networks. Significant improvement in throughput and signal-to-interference-and-noise-ratio ($\mathtt{SINR}$) have been shown in \cite{ElsBoc14} with realistic simulations, while \cite{SmiPop15, BooBha15, SinXha14} reached similar conclusions from a theoretical perspective.  In particular, the key idea is that in many cases the uplink throughput can be improved significantly by connecting to a small cell in the UL while being on a macrocell in the DL.  A recent survey of these results, with a discussion of how to adapt it to 4G and 5G cellular standards, is given in \cite{BocAnd15}.  


Meanwhile, starting with \cite{AndBac11}, modeling and analyzing cellular networks using stochastic geometry has become a popular and accepted approach to understanding their performance trends. Most relevant to this study, mmWave networks were analyzed assuming a Poisson point process (PPP) for the base station (BS) distribution in \cite{BaiHea14,SinKul14, ParKim15}. In \cite{BaiHea14} a line-of-sight (LOS) ball model was considered for blockage modeling where BSs inside the LOS ball were considered to be in LOS whereas any BS outside of the LOS ball was treated as NLOS.  In \cite{SinKul14}, this blocking model was modified by adding a LOS probability within the LOS ball, and this approach was shown to reflect several realistic blockage scenarios. Therefore we consider the same approach in this paper. Decoupled association in a mixed sub-6GHz and mmWave deployment was very recently considered in \cite{ParKim15} from a resource allocation perspective. However, there is no complete or analytical study to our knowledge on downlink-uplink decoupling for mmWave networks or the mmWave-sub-6GHz hybrid network considered in this paper.

\subsection{Contributions and Organization}

 In Section \ref{system_model}, we model  a cellular network with sub-6GHz macrocells (Mcells) and mmWave small cells (Scells) each distributed according to an independent Poisson point process.  A UE can in general independently connect to either type of BS on the UL and DL.  The key technical contributions of this paper are the following.

\textbf{Cell association probabilities}. In Section \ref{CA_derivation}, we derive  the cell association probabilities based on the UL and DL maximum biased  received power where the different parameters that affect the association trends are highlighted and discussed in detail. Subsequently, a similar analysis based on the UL and DL maximum achievable rate is given.  The role of decoupled access is discussed in detail in Section \ref{validation}.

\textbf{Coverage and rate trends}. The UL and DL $\mathtt{SINR}$ and rate coverage probabilities are derived in Section \ref{SINRratederv} where a special emphasis is put on Scell biasing.  We show that high biasing values can be used for mmWave Scells due to the abundant bandwidth in the mmWave bands. The altered UL and DL $\mathtt{SINR}$ and rate coverage with the biasing value are also studied.

\textbf{System design insights}. The analytical results, which employ a number of simplifying approximations, are validated in Section \ref{validation}.  Design insights are highlighted in Section \ref{design_implications} which include:
\begin{itemize}
\item Decoupled access plays a key role in mmWave deployments and the gains of decoupling are more pronounced in less dense urban environments.
\item Scell beamforming gain improves the association probability to Scells dramatically and therefore needs to be considered in the association phase.
\item Aggressive values of small cell biasing are possible thanks to the wide bandwidth offered by mmWaves. Supporting these large biasing values requires having robust low modulation and coding techniques to allow UEs to operate in very low $\mathtt{SINR}$.
\end{itemize}



\section{System model}
\label{system_model}
\subsection{Spatial distributions}

A two-tier heterogeneous network is considered where Mcells and Scells are distributed uniformly in $\mathbb{R}^2$ according to independent homogeneous Poisson point processes (PPP) $\Phi_m$ and $\Phi_s$ with densities $\lambda_m$ and $\lambda_s$ respectively. Specifically, a deployment of sub-6GHz Mcells overlaid by mmWave Scells is considered. The UEs are also assumed to be uniformly distributed according to a homogeneous PPP  $\Phi_u$  with density $\lambda_u$.
The analysis is done for a typical UE located at the origin where the BS serving the typical UE is referred to as the tagged BS \footnote{The analysis of the typical UE is enabled by Slivnyak's theorem.}. The notation is summarized in Table \ref{tab:Notation_1}. The inclusion of sub-6GHz Scells is left for future work.

\begin{table}[!t]
\caption{Notation and simulation parameters}
\centering
\begin{tabular}{ | m{5em} | m{6cm}| m{3cm} | }
\hline
\textbf{Notation} & \textbf{Parameter} & \textbf{Value (if applicable)}\\
\hline
$\Phi_m$, $\lambda_m$ & Mcells PPP and density & $\lambda_m$ = 5 per sq. km\\
\hline
$\Phi_s$, $\lambda_s$ & Scells PPP and density &$\lambda_s$ = 50 per sq. km\\
\hline
$\Phi_u$, $\lambda_u$ & UEs PPP and density & $\lambda_u$ = 200 per sq. km\\
\hline
$\mathrm{f}_{m}$, $\mathrm{f}_{s}$  & sub-6GHz and mmWave carrier frequencies & 2 GHz, 70 GHz \\
\hline
$\mathrm{W}_m$, $\mathrm{W}_s$ & sub-6GHz, mmWave bandwidth &20 MHz, 1 GHz\\
\hline
$\mathrm{P}_m$, $\mathrm{\mathrm{P}_s}$ & Mcell and Scell transmit power &46 dBm, 30 dBm\\
\hline
$\mathrm{P}_{um}$, $\mathrm{P}_{us}$ & UE transmit power to Mcell and Scell &23 dBm\\
\hline
$K_{\mathrm{UL}}$, $K_{\mathrm{DL}}$ & UL and DL association tiers & \\
\hline
$\mathrm{T}_{s}$, $\mathrm{T}_{s}'$ & DL and UL association bias of mmWave Scells & \\
\hline
$\mathrm{T}_{m}$, $\mathrm{T}_{m}'$ & DL and UL association bias of sub-6GHz Mcells & \\
\hline
$\alpha_m$ & Pathloss exponent for Mcells & 3 \\
\hline
$\alpha_{l}$, $\alpha_{n}$ & LOS and NLOS pathloss exponent for Scells & 2, 4 \\
\hline
$G_{s_{max}}$, $G_{s_{min}}$, $\theta_s$ & Main lobe gain, side lobe gain and 3 dB beamwidth for mmWave & 18 dBi, -2 dBi, 10$\degree$ \\
\hline
$G_m$ & Mcell antenna gain (omni-directional) & 0 dBi  \\
\hline
$\mathrm{\omega}$, $\mathrm{\mu}$ & Fractional LOS area $\mathrm{\omega}$ in a ball of radius $\mathrm{\mu}$ & 0.11, 200 m \\
\hline
$N_m$, $N_s$ & Load of serving Macro or Small cell &  \\
\hline
$\mathcal{A}$, $\mathcal{B}$ & Association probabilities based on max. biased received power and max. rate &  \\
\hline
$h$ &Small scale fading& $h \sim \exp(1)$ \\
\hline
$\beta$ & $\beta=\left(\frac{\mathrm{carrier \ wavelength}}{4\pi}\right)^2$ is the pathloss at 1m&  \\
\hline
$\sigma_m^2$, $\sigma_s^2$ & Noise powers for sub-6GHz and mmWave & -174 dBm/Hz + 10$\log_{10}(\mathrm{W})$ + 10 dB\\
\hline

\end{tabular}
\label{tab:Notation_1}
\end{table}

\subsection{Propagation assumptions}
The received power in the DL at a UE at location $u\in\Phi_{u}$ from a sub-6GHz Mcell ($m$) at $x \in \Phi_m $ or a mmWave Scell ($s$) at $y \in \Phi_s $  is given by $\mathrm{P}_m h_{x,u}\beta_m G_m L_{m}(x-u)^{-1}$ or $\mathrm{P}_sh_{y,u}\beta_s G_s(\theta) L_{s}(y-u)^{-1}$, respectively. Here, $L$ is the pathloss where for the typical UE at the origin $L_{m}(x) = \left\|x\right\| ^ {\alpha_m}$ and $L_{s}(y) = \left\|y\right\| ^ {\alpha_s(y)}$, $\alpha$ is the pathloss exponent (PLE) where $\alpha_s(y)$ equals $\alpha_l$ if the link is LOS and $\alpha_n$ otherwise, $h$ is the small scale fading power gain where in this study we consider Rayleigh fading, $\beta$ is the the near-field pathloss at 1 m and $G$ is the antenna gain. UEs are assumed to have omni-directional antennas so the antenna gains are only accounted for at the BS side. All mmWave Scells are equipped with directional antennas with a sectorized gain pattern assuming a simplified rectangular antenna pattern that was used in \cite{SinKul14} where a UE receives a signal with $G_{s_{max}}$ if the UE's angle ($\theta$) with respect to the best beam alignment is within the main beamwidth $(\theta_s)$ of the serving cell and $G_{s_{min}}$ otherwise. This is formulated by
 \[
 G_s(\theta) = \begin{cases}
 G_{s_{max}} \hspace{1em} if \ |\theta| \leq \ \frac{\theta_s}{2} \\
 G_{s_{min}} \hspace{1em} otherwise
 \end{cases}.
\]
The UL received signal powers are derived by replacing $\mathrm{P}_m$ or $\mathrm{P}_s$ by $\mathrm{P}_{um}$ or $\mathrm{P}_{us}$, and interchanging $x$ or $y$ with $u$, respectively. Shadowing is ignored in this study since for mmWaves the blockage model introduces a similar effect to shadowing. As for the sub-6GHz network, as shown in \cite{AndBac11},  the randomness of the PPP BS locations emulates the shadowing effect, therefore shadowing is ignored in the sub-6GHz model as well.

All UEs served by the Scells are assumed to be in perfect alignment with their serving cells whereas the beams of all interfering links are assumed to be randomly oriented with respect to each other and hence the gain on the interfering links is considered to be random. In  Section~\ref{Assoc_results1}  results for the association probabilities considering different antenna gains are shown in order to study how important it is to have antenna alignment in the cell association phase.

\subsection{Blockage model}
\label{blockage_model}
A simple yet accurate blockage model that was proposed in \cite{SinKul14} is used where a UE within a distance $\mathrm{\mu}$ from a Scell is assumed LOS with probability $\mathrm{\omega}$ and 0 otherwise. The parameters $\mathrm{\omega}$ and $\mathrm{\mu}$ are environment dependent; the Manhattan scenario from \cite{SinKul14} is considered for this study. Results for other values of $\mathrm{\omega}$ and $\mathrm{\mu}$ are shown in Section~\ref{Assoc_results1} to study their effect on cell association.

\subsection{Biased uplink and downlink cell association}

 

It is assumed that the UL and DL cell associations are based on different criteria, namely the UL and DL biased received powers, respectively. The typical user associates with BS at $x^*\in \Phi_{l}$, where $l\in \{s,m\}$, in UL if and only if 
 \begin{eqnarray}
\mathrm{P}_{ul} \mathrm{T}'_{l} \psi_l L_{l}(x^*)^{-1} \geq \mathrm{P}_{uk} \mathrm{T}'_{k} \psi_k L_{\min ,k}^{-1}, \ \forall  k  \in \{s,m\},
 \end{eqnarray}
where $\psi_k = G_k \beta_k$ is the combination of antenna gain and near-field pathloss and $G_k$ is equal to $G_{s_{max}}$ or $G_m$ in the mmWave or sub-6GHz cases, respectively. $ L_{\min ,k} = \min_{x \in \Phi_k} L_k(x)$ is the minimum pathloss of the typical UE from the $k^{th}$ tier and $ \mathrm{T}'$ and  $\mathrm{T}$ are the UL and DL cell bias values respectively. Similarly, the typical user associates with BS at $x^*\in \Phi_{l}$ in DL if and only if 
\begin{eqnarray}
\mathrm{P}_{l} \mathrm{T}_{l} \psi_l L_{l}(x^*)^{-1} \geq \mathrm{P}_{k} \mathrm{T}_{k} \psi_k L_{\min ,k}^{-1}, \ \forall  k  \in  \{s,m\}.
\end{eqnarray}

The assumption that large bandwidth mmWave networks are noise-limited has been considered and motivated in \cite{SinKul14}. We show in Section \ref{validation} that this assumption holds even for high densities of mmWave Scells.  Henceforth, this assumption will be considered for this study and is validated later on with simulation results.
Consequently and in order to simplify the analysis, the signal-to-noise-ratio ($\mathtt{SNR}$) is considered instead of the $\mathtt{SINR}$ for the mmWave links. With no interference between the two tiers due to the orthogonality of both frequency bands, the UL/DL sub-6GHz $\mathtt{SINR}$ and mmWave $\mathtt{SNR}$ of a typical UE at the origin  are given by
\begin{eqnarray*}
\mathtt{SINR}_{\mathrm{UL},m} = \frac{\mathrm{P}_{um} \psi_{m} h_{0,x^*} L_{m}(x^*)^{-1}}{ I_{\mathrm{UL},m} + \sigma_m^2}, \ \
\mathtt{SINR}_{\mathrm{DL},m} = \frac{\mathrm{P}_{m} \psi_{m}h_{x^*,0} L_{m}(x^*)^{-1}}{I_{\mathrm{DL},m} + \sigma_m^2},
\end{eqnarray*}
\begin{eqnarray}
\mathtt{SNR}_{\mathrm{UL},s} =  \frac{\mathrm{P}_{us} \psi_{s}h_{0,x^*} L_{s}(x^*)^{-1}}{\sigma_s^2}, \ \ \mathtt{SNR}_{\mathrm{DL},s} =  \frac{\mathrm{P}_{s} \psi_{s}h_{x^*,0} L_{s}(x^*)^{-1}}{\sigma_s^2},
\end{eqnarray}
where $I_{\mathrm{UL},m} = \sum\limits_{{y} \in {\Phi_{Iu}}} \mathrm{P}_{um} \psi_{m}h_{y,x^*}L_m(y-x^*)^{-1}$, $I_{\mathrm{DL},m} = \sum\limits_{{x} \in {\Phi_{m}}\backslash x^*} \mathrm{P}_{m} \psi_{m}h_{x,0} L_{m}(x)^{-1}$ and $\Phi_{Iu}$ is the point process denoting the locations of UEs transmitting in the UL on the same resource as the typical UE.  It is assumed that each BS has at least one UE in its association region. With this assumption, the realizations of $\Phi_{Iu}$ have one point randomly chosen from the association cell of each BS other than the serving BS, which represents  the interfering UE ($y$) from that cell in the UL. Furthermore, the queues in the UL and DL are assumed to be always full  and resources are on average equally distributed among the UEs (e.g. by proportional fair or round robin scheduling). 
The DL rate of the typical UE connected to a Mcell or Scell is given by
\begin{eqnarray}
 R_{\mathrm{DL},m} = \frac{\mathrm{W}_m}{N_m} \log(1+\mathtt{SINR}_{\mathrm{DL},m}), \ \  R_{\mathrm{DL},s} = \frac{\mathrm{W}_s}{N_s} \log(1+\mathtt{SNR}_{\mathrm{DL},s}),
  \end{eqnarray}
  where $N_m$ and $N_s$ are the loads on the serving Mcell and Scell respectively. $R_{\mathrm{UL},m}, R_{\mathrm{UL},s}$ are defined similarly. 

  \section{Cell association}
	\label{CA_derivation}
  In this section the UL and DL cell association probabilities are derived for four different cases where $K_{\mathrm{DL}}$ and $K_{\mathrm{UL}}$ denote the DL and UL association tiers of the typical UE.  Hence, the below cases denote the probability of the UE  associating to the Mcell and Scell in the UL and DL assuming a decoupled UL and DL association approach.
  \begin{itemize}
   \item  Case 1: $\mathbb{P}(K_{\mathrm{DL}} = Mcell)$
  \item Case 2: $\mathbb{P}(K_{\mathrm{UL}} = Mcell)$ 
  \item Case 3: $\mathbb{P}(K_{\mathrm{DL}} = Scell)$ 
  \item Case 4: $\mathbb{P}(K_{\mathrm{UL}} = Scell)$ 
\end{itemize}
 
 Note that the sum of probabilities of Case 1 and 3 equals 1 and similarly for Case 2 and 4.
 The association probabilities are derived in the following two subsections, maximizing the biased DL/UL received power and the DL/UL rate, respectively. Subsequently, the outcomes from the two association strategies are compared in Section \ref{Assoc_results1}.

In order to derive the association probabilities, we first characterize the point process formed by the pathloss between each BS and the typical UE at the origin. Assuming a BS at $x \in \mathbb{R}^2$, the pathloss point process is defined as $\mathcal{N}_l :=\{ L_l(x) = \|x\|^{\alpha_l}  \}_{x \in \Phi_l}$, where $l\in \{m,s\}$. Making use of the displacement theorem, $\mathcal{N}_l$  is a Poisson point process with intensity measure denoted by $\Lambda_l(.)$ similar to \cite{SinKul14 , BlaKar13}. Since the pathloss in the sub-6GHz and mmWave cases has different characteristics, we will have two independent pathloss processes for mmWave and sub-6GHz given by  $\mathcal{N}_s$ and $\mathcal{N}_m$ respectively. Therefore, the intensities, probability distribution function (PDF) and complementary cumulative distribution function (CCDF) will be derived separately for mmWave and sub-6GHz.

\textbf{ Lemma 1.} \textit{ The distribution of the pathloss from the typical UE to the tagged BS is such that} $\mathbb{P}(L_l(x) > t) = \exp(- \Lambda_l((0,t)])$, \textit{where} $l\in \{m,s\}$,  \textit{the intensity measures for pathloss in mmWave and  sub-6GHz are given by}
\begin{align} \label{lambda_s}
\Lambda_s((0,t)] &= \pi \lambda_s \Bigg( \left(\mathrm{\omega} t^{\frac{2}{\alpha_l}} + (1-\mathrm{\omega})t^{\frac{2}{\alpha_n}}\right)\mathds{1}(t < \mathrm{\mu}^{\alpha_l})+  \left(\mathrm{\omega} \mathrm{\mu}^2 + (1-\mathrm{\omega}) t^{\frac{2}{\alpha_n}} \right)  \mathds{1}( \mathrm{\mu}^{\alpha_l} \leq t \leq \mathrm{\mu}^{\alpha_n})\nonumber \\ & + t^{\frac{2}{\alpha_n}} \mathds{1} (t > \mathrm{\mu}^{\alpha_n})\Bigg)\\
\Lambda_m((0,t)] &= \pi \lambda_m t^{\frac{2}{\alpha_m}}.\label{lambda_m}
\end{align}


\begin{proof}[Proof:\nopunct] See Appendix \ref{lemma1_proof}.\end{proof}

Since  $\mathcal{N}_l$ is a PPP, the CCDF of pathloss to the tagged BS is $ \bar{F}_l(t) = \mathbb{P} (L_l(x) > t) = \exp(- \Lambda_l((0,t]))$
 and the PDF is given by $f_l(t) = \frac{-d \bar{F}_l(t) }{dt}  =\Lambda_l'((0,t]) \exp( - \Lambda_l((0,t]))$ for $l \in (m,s)$.
The expressions for the pathloss process CCDFs for mmWave  and sub-6GHz are given by 
\begin{align}
\bar{F}_s(t) &=  \exp\Bigg(-\pi \lambda_s \Big( \left(\mathrm{\omega} t^{\frac{2}{\alpha_l}} + (1-\mathrm{\omega})t^{\frac{2}{\alpha_n}}\right)\mathds{1}(t < \mathrm{\mu}^{\alpha_l})+  \left(\mathrm{\omega} \mathrm{\mu}^2 + (1-\mathrm{\omega}) t^{\frac{2}{\alpha_n}} \right)\mathds{1}( \mathrm{\mu}^{\alpha_l} \leq t \leq \mathrm{\mu}^{\alpha_n}) \nonumber \\ &+ t^{\frac{2}{\alpha_n}} \mathds{1} (t > \mathrm{\mu}^{\alpha_n})\Big)\Bigg)\label{fcap_s}\\
\bar{F}_m(t) &= \exp\left(-\pi \lambda_m t^{\frac{2}{\alpha_m}}\right)
\end{align}
and the corresponding PDFs by
\begin{align}
f_s(t) &= 2 \pi \lambda_s \frac{t^{\frac{2}{\alpha_n}- 1}}{\alpha_n} \Bigg( \bigg(\frac{\alpha_n \mathrm{\omega} t^{\frac{2}{\alpha_l} - \frac{2}{\alpha_n}}}{\alpha_l}  + (1-\mathrm{\omega})\bigg)  \exp\left(-\pi \lambda_s \left( \mathrm{\omega} t^{\frac{2}{\alpha_l}} + (1-\mathrm{\omega})t^{\frac{2}{\alpha_n}}\right)\right)\mathds{1}(t < \mathrm{\mu}^{\alpha_l}) \nonumber \\ &+ (1-\mathrm{\omega}) \exp\left(-\pi \lambda_s \left( \mathrm{\omega} \mathrm{\mu}^2 + (1-\mathrm{\omega}) t^{\frac{2}{\alpha_n}}\right) \right)\mathds{1}( \mathrm{\mu}^{\alpha_l} \leq t \leq \mathrm{\mu}^{\alpha_n}) + \exp\left(-\pi \lambda_s t^{\frac{2}{\alpha_n}}\right) \mathds{1} (t > \mathrm{\mu}^{\alpha_n})\Bigg)\label{f_s}\\
f_m(t) &= \frac{2 \pi \lambda_m t^{\frac{2}{\alpha_m}-1}}{\alpha_m} \exp\left(\pi \lambda_m t^{\frac{2}{\alpha_m}}\right).\label{f_m}
\end{align}
All the needed components to derive the association probabilities specified above are now available.
\subsection{Maximum biased received power association}
\label{Max_RP}
 In this subsection, the UL and DL association probabilities maximizing the biased UL and DL received power respectively are derived. This method is referred to as maximum biased received power (Max-BRP). It is assumed that the DL and UL serving cells are chosen based on the biased DL and UL  received powers respectively. The association probabilities are defined in the following definition and the final expressions are given in Lemma 2.

\textbf{Definition 1. Max-BRP Association probabilities.} The probabilities of the typical UE associating to a sub-6GHz  Mcell or mmWave Scell based on the maximum biased received power in the downlink or uplink is defined as
\begin{align}
\mathcal{A}_{\mathrm{DL}, m} \triangleq \mathbb{P} \left( \mathrm{P}_{m} \mathrm{T}_{m} \psi_m L_{\min,m}^{-1} > \mathrm{P_{s}} \mathrm{T}_{s} \psi_s L_{\min,s}^{-1} \right)
\end{align}
\begin{align}
 \mathcal{A}_{\mathrm{UL}, m} \triangleq \mathbb{P} \left( \mathrm{P}_{um} \mathrm{T}_{m}' \psi_m L_{\min,m}^{-1} > \mathrm{P}_{us} \mathrm{T}_{s}' \psi_s L_{\min,s}^{-1} \right) 
\end{align}
\begin{align}
 \mathcal{A}_{\mathrm{DL}, s} \triangleq \mathbb{P} \left( \mathrm{P_{s}} \mathrm{T}_{s} \psi_s L_{\min,s}^{-1}  > \mathrm{P}_{m} \mathrm{T}_{m} \psi_m L_{\min,m}^{-1}   \right)
\end{align}
\begin{align}
\mathcal{A}_{\mathrm{UL}, s} \triangleq \mathbb{P} \left( \mathrm{P}_{us} \mathrm{T}_{s}' \psi_s L_{\min,s}^{-1}  > \mathrm{P}_{um} \mathrm{T}_{m}' \psi_m L_{\min,m}^{-1}   \right).
\end{align}
\textbf{Lemma 2.} \textit{The uplink and downlink association probability to a sub-6GHz Mcell or mmWave Scell are given below.}
\begin{eqnarray}
\begin{aligned}
\mathcal{A}_{c, m} &= \frac{2 \pi \lambda_m}{\alpha_m a_c^{\frac{2}{\alpha_m}}} \int\limits_0^{\infty} l^{\frac{2}{\alpha_m }-1} \exp\Bigg( - \pi \lambda_m \left(\frac{l}{a_c}\right)^{\frac{2}{\alpha_m}}\Bigg)\Bigg( \exp\left(-\pi\lambda_s (\mathrm{\omega} l^{\frac{2}{\alpha_l}} + (1-\mathrm{\omega})l^{\frac{2}{\alpha_n}} ) \right)\mathds{1}(l < \mathrm{\mu}^{\alpha_l}) \\ & +  \exp\left( -\pi \lambda_s \left((1-\mathrm{\omega}) l^{\frac{2}{\alpha_n}} +\mathrm{\omega} \mathrm{\mu}^2\right) \right)\mathds{1}( \mathrm{\mu}^{\alpha_l} \leq l \leq \mathrm{\mu}^{\alpha_n}) + \exp\left( -\pi\lambda_s  l^{\frac{2}{\alpha_n}}\right)\mathds{1} (l > \mathrm{\mu}^{\alpha_n}) \Bigg) \mathrm{d}l \label{A_M_DL}\\
\mathcal{A}_{c, s} &= 1 - \mathcal{A}_{c, m},
\end{aligned}
\end{eqnarray}

where  $c \in \{\mathrm{UL}, \mathrm{DL}\}$, $ a_{\mathrm{DL}} = \frac{\mathrm{P}_{s} \mathrm{T}_{s} \psi_s}{\mathrm{P}_{m} \mathrm{T}_{m} \psi_m}$ and $a_{\mathrm{UL}} = \frac{\mathrm{P}_{us} \mathrm{T}_{s}' \psi_s}{\mathrm{P}_{um} \mathrm{T}_{m}' \psi_m}$. \\

\begin{proof}[Proof:\nopunct]
The proof for $\mathcal{A}_{\mathrm{DL}, m}$ is given below.
\begin{eqnarray*}
\mathcal{A}_{\mathrm{DL}, m} &=& \mathbb{P} \left( \mathrm{P}_{um} \mathrm{T}_{m} \psi_m L_{min,m}^{-1} > \mathrm{P}_{us} \mathrm{T}_{s} \psi_s L_{min,s}^{-1} \right) = \mathbb{P} \left(L_{min,s} > a_{\mathrm{DL}} L_{min,m} \right)  \\
&=&  \int \limits_0^{\infty}  \bar{F}_{s}(a_{\mathrm{DL}} l_m) f_{m}(l_m) \mathrm{d}l_m,
\end{eqnarray*}

where the last step follows from the fact that  $\mathbb{P} (X > Y) = \int\limits_0^{\infty} \mathbb{P} (X>y) f_Y(y) \mathrm{d}y $.

Changing variables as  $ l = a_{\mathrm{DL}} l_m $ yields
\begin{eqnarray*}
\mathcal{A}_{\mathrm{DL}, m}  = \frac{1}{a_{\mathrm{DL}}} \int\limits_0^{\infty}  \bar{F}_{s}(l) f_{m}\left(\frac{l}{a_{\mathrm{DL}}} \right) \mathrm{d}l.
\end{eqnarray*}
This directly results in $\mathcal{A}_{\mathrm{DL}, m}$, and $\mathcal{A}_{\mathrm{UL}, m}$ follows similarly.
\end{proof}
\textbf{Corollary 1.} \textit{The association probabilities can be acquired in closed form for the special case where $\alpha_l = 2$ and $\alpha_n = \alpha_m = 4$ and with simple mathematical manipulation, $\mathcal{A}_{\mathrm{DL}, m}$ can be expressed by}
\begin{eqnarray}
\begin{aligned}
\mathcal{A}_{ \mathrm{DL},m} = \frac{\pi \lambda_m}{\sqrt{a_{\mathrm{DL}}}} \Bigg( \frac{\sqrt{\pi} e^{\frac{c_2^2}{4 c_1}}}{2\sqrt{c_1}}\left(Q\left(\frac{c_2}{\sqrt{2 c_1}}\right) - Q \left(\frac{2 \mathrm{\mu} c_1 + c_2}{\sqrt{2 c_1}}\right)\right) + e^{-\mathrm{\mu}^2 c_1} \left( \frac{e^{-\mathrm{\mu} c_2}}{c_2} - \frac{c_1 e^{-\mathrm{\mu}^2 c_2}}{c_2(c_1 + c_2)} \right)\Bigg),
\end{aligned}
\end{eqnarray}
where $c_1 = \pi \lambda_s \mathrm{\omega} $, $c_2 = \pi \lambda_s (1 - \mathrm{\omega}) + \frac{\pi \lambda_m}{a_{\mathrm{DL}}^{2/\alpha_n}} $ and $Q(.)$ is the Q-function.
Similarly, the other three cases can be obtained in closed form.

\subsection{Max-Rate association}
\label{Max_rate_sec}
In this part the UL and DL association probabilities are derived where the association criteria are the UL and DL rates respectively. The sub-6GHz Macro DL association probability is given by
\begin{align}
\mathcal{B}_{\mathrm{DL}, m} = \mathbb{P} \left( \frac{\mathrm{W}_m}{N_{\mathrm{DL} ,m}} \log_2(1+\mathtt{SINR}_{\mathrm{DL},m}) > \frac{\mathrm{W}_s}{N_{\mathrm{DL},s}} \log_2(1+\mathtt{SINR}_{\mathrm{DL},s}) \right).\nonumber
\end{align}
It is assumed that $\mathtt{SINR}_{\mathrm{DL},m}  \approx \mathtt{SIR}_{\mathrm{DL},m}$ and $\mathtt{SINR}_{\mathrm{DL},s}  \approx \mathtt{SNR}_{\mathrm{DL},s}$ for simplicity since sub-6GHz frequencies are interference limited whereas mmWaves are rather noise limited.
In order to simplify the expressions, an approximation\footnote{This approximation was proposed for sub-6GHz in \cite{SinAnd13} and was later verified for mmWaves in \cite{SinKul14}.} that was proposed in \cite{SinAnd13} is used where the cell load is characterized by the average number of UEs per cell on the corresponding tier. The average load on the serving BS of tier $l$ for the UL and DL is given by
\begin{eqnarray}
\bar{N}_{c,l} = 1 + \frac{1.28 \lambda_u \mathcal{B}_{c,l}}{\lambda_c} \ \ \mathrm{for} \ l \in \{m,s\}\ \mathrm{and}\ c \in \{\mathrm{UL},\mathrm{DL}\}.\label{load_approx}
\end{eqnarray}


This approximation results in 
\begin{align}
\mathcal{B}_{\mathrm{DL}, m} = \mathbb{P} \left(  \mathtt{SIR}_{\mathrm{DL},m} > (1+\mathtt{SNR}_{\mathrm{DL},s})^{\left(\frac{\mathrm{W}_{s}(\lambda_m + 1.28 \lambda_u \mathcal{B}_{\mathrm{DL}, m}) \lambda_s}{\mathrm{W}_{m}(\lambda_s + 1.28 \lambda_u \mathcal{B}_{\mathrm{DL}, s})\lambda_m}  \right)} -1 \right).\label{assoc_rate1}
\end{align}
Having $\mathcal{B}_{\mathrm{DL}, m} $ on both sides of the equation makes it very hard to solve. Therefore we resort to a simple approximation by neglecting the load term in the rate expression (setting $N_m$ and $N_s$ to 1). In other words deriving the association probability based on the maximum achievable rate in the UL and DL. This approach is suboptimal but it results in a tractable expression for the association probability and also suffices our purpose of showing different decoupling trends as compared to Max-BRP as will be shown in Section \ref{validation}. The association trends resulting from this approximation are also validated in Fig. \ref{fig:SINR_rate_sim1}(b). 
Henceforth this method is referred to as \textit{Max-Rate} and the corresponding association probabilities are now defined.

\textbf{Definition 2. Max-Rate Association Probability.}  The association probabilities in the UL and DL to a sub-6GHz Mcells and mmWave Scells for the Max-Rate case are given by 
\begin{align}
\mathcal{B}_{c, m} \triangleq \mathbb{P} \left(  \mathtt{SIR}_{c,m} > (1+\mathtt{SNR}_{c,s})^{\left(\frac{\mathrm{W}_{s}}{\mathrm{W}_{m}}  \right)} -1 \right)\label{assoc_rate2}
\end{align}
\begin{align}
\mathcal{B}_{c, s} \triangleq \mathbb{P} \left(  \mathtt{SNR}_{c,s} > (1+\mathtt{SIR}_{c,m})^{\left(\frac{\mathrm{W}_{m}}{\mathrm{W}_{s}}  \right)} -1 \right),\label{assoc_rate2}
\end{align}
where $c \in \{\mathrm{UL}, \mathrm{DL}\}$.

Using power control in the UL would  complicate the UL coverage expression where the derived expressions in \cite{NovDhi13} and \cite{SinXha14} include two to three integrals. Furthermore, it is assumed that UEs transmit with their maximum power on mmWaves since mmWaves are coverage limited, therefore to make the analysis consistent and fair we assume that UEs transmit with their maximum power on sub-6GHz as well. 
With the assumption of no power control the UL and DL coverage expressions (neglecting noise and considering exponential fading) are the same. In order for this assumption to be valid we also need to assume that the interferers in the UL are PPP distributed and that the exclusion region around the typical UE/BS in the DL/UL are the same. Although the latter might seem to be a strong assumption it will be shown in Fig. \ref{fig:AssocProb_3_main}(\subref{fig:AsspcoProb_r}) that the derived rate based association probability matches very well the simulation results, verifying that the above assumptions are valid.  

The final expressions for the Max-Rate based association probabilities based on Definition 2 are given in the following lemma.

\textbf{Lemma 3.} \textit{ The DL and UL association probability based on the maximum achievable rate for mmWave Scells and sub-6GHz Mcells are given below}. \\
\begin{align}
\mathcal{B}_{c, m} &=  \int\limits_0^{\infty}\frac{f_{\mathtt{SNR}_{c,s}}(z)}{1+ \rho\left((1+ z)^{\frac{\mathrm{W}_{s}}{\mathrm{W}_{m}}} -1 \right)} \mathrm{d}z \\
\mathcal{B}_{c, s} &= 1 - \mathcal{B}_{c, m},
\end{align}
where $c \in \{\mathrm{UL}, \mathrm{DL}\}$, $f_{\mathtt{SNR}_{\mathrm{DL},s}(z)} = \frac{\sigma_s^2}{\mathrm{P}_s \psi_s} \int\limits_0^\infty l \exp\left(\frac{- z \sigma_s^2 l}{\mathrm{P_s \psi_s}}  \right) f_s(l) dl$, $\rho (t, \alpha_m) = t^{\frac{2}{\alpha_m}} \int\limits_{t^{\frac{-2}{\alpha_m}}}^{\infty} \frac{\mathrm{d}u}{1 + u^{\frac{\alpha_m}{2}}} $ and  $f_{\mathtt{SNR}_{\mathrm{UL},s}}$ is the same as $f_{\mathtt{SNR}_{\mathrm{DL},s}}$ exchanging $\mathrm{P}_s$ by $\mathrm{P}_{us}$ .

\begin{proof}[Proof:\nopunct] See Appendix \ref{lemma3_proof}. \end{proof}

After deriving the association probabilities, the UL and DL $\mathtt{SINR}$ and rate coverage probabilities are derived in the next section.

\section{SINR and rate distributions: Downlink and Uplink}
\label{SINRratederv}
In this section the $\mathtt{SINR}$ and rate coverage distributions are derived for the DL and UL in the mmWave and sub-6GHz cases. These distributions would help in studying the effect of the different  association strategies on the $\mathtt{SINR}$  and rate of the whole system.
The $\mathtt{SINR}$ and rate CCDFs will be derived for the Max-BRP association case only as the derivation for the Max-Rate association is quite complicated and will be left for future work. However, we use biasing in the results for $\mathtt{SINR}$ and rate coverage probabilities  in Section \ref{validation} to validate some of the trends that result from the Max-Rate association strategy.

\subsection{SINR coverage}
 The $\mathtt{SINR}$  coverage can be defined as the average fraction of UEs that at any given time achieve $\mathtt{SINR}$  $\tau$. The $\mathtt{SINR}$  coverage is the CCDF of the $\mathtt{SINR}$  over the entire network which, due to the assumption of stationary PPP for the UEs and BSs, can be characterized considering the typical link between the typical UE at the origin and its serving BS. Since mmWave networks are usually noise limited (i.e. $ \mathtt{SINR}  \approx \mathtt{SNR} $), we consider the $\mathtt{SNR}$ coverage for mmWaves while still considering $\mathtt{SINR}$ for sub-6GHz. The $\mathtt{SINR}/\mathtt{SNR} $ coverage in the sub-6GHz and mmWave cases is expressed as:  $\mathcal{P}_m  \triangleq \mathbb{P}(\mathtt{SINR}  >\tau)  $ and $\mathcal{P}_s  \triangleq \mathbb{P}(\mathtt{SNR}  >\tau)  $ respectively.
 Since there is no interference between the mmWave and sub-6GHz BSs, the $\mathtt{SINR}/\mathtt{SNR}$  coverage can be derived separately for  sub-6GHz and mmWave.
 
 Similar to Section \ref{Max_rate_sec}, since  UL transmissions on mmWaves are assumed to be at maximum power,  the sub-6GHz UL $\mathtt{SINR}$  coverage is derived assuming maximum UL transmit power (no power control) for simplicity and fairness. The final expression for the UL coverage probability with fractional pathloss compensation power control is given below.

 \textbf{Theorem 1.} \textit{The $\mathtt{SINR}$  coverage probability for the typical UL and DL links based on the Max-BRP association criterion is given by}
\begin{eqnarray}
\begin{aligned}
\mathcal{P}_{\mathrm{DL}} (\tau) &= \mathcal{P}_{\mathrm{DL} , m} (\tau) + \mathcal{P}_{\mathrm{DL}, s} (\tau)\\
&= \int \limits_0^{\infty} \exp\left( \frac{-\tau \sigma_m^2 l}{\mathrm{P}_{m}  \psi_m} \right) \exp\left( \frac{-2\pi \lambda_m}{\alpha_m} \int \limits_l^{\infty} \frac{t^{\frac{2}{\alpha_m}-1}}{1 + \frac{t}{\tau l}} \mathrm{d}t \right) \bar{F}_s (a_{\mathrm{DL}} l) f_m(l) \mathrm{d}l\\ &+ \int \limits_0^{\infty}\exp\left( \frac{-\tau \sigma_s^2 l}{\mathrm{P_{s}}  \psi_s} \right)  \bar{F}_m \left(\frac{l}{a_{DL}}\right) f_s(l) \mathrm{d}l
\end{aligned}
\end{eqnarray}

\begin{eqnarray}
\begin{aligned}
\mathcal{P}_{\mathrm{UL}} (\tau) &= \mathcal{P}_{\mathrm{UL} , m} (\tau) + \mathcal{P}_{\mathrm{UL} , s} (\tau)\\
&= \int \limits_0^{\infty} \exp\left( \frac{-\tau \sigma_m^2 l}{\mathrm{P}_{um}  \psi_m} \right) \exp\left( \frac{-2\pi \lambda_m}{\alpha_m} \int \limits_l^{\infty} \frac{t^{\frac{2}{\alpha_m}-1}}{1 + \frac{t}{\tau l}} \mathrm{d}t \right) \bar{F}_s (a_{\mathrm{UL}} l) f_m(l) \mathrm{d}l\\ &+\int \limits_0^{\infty} \exp\left( \frac{-\tau \sigma_s^2 l}{\mathrm{P}_{us}  \psi_s} \right)  \bar{F}_m \left(\frac{l}{a_{\mathrm{UL}}}\right) f_s(l) \mathrm{d}l,
\end{aligned}
\end{eqnarray}

 where $\bar{F_s}, \bar{F_m}, f_s$, $f_m$, $a_{\mathrm{DL}}$ and $a_{\mathrm{UL}}$ have been derived/defined in Section \ref{CA_derivation}.

\begin{proof}[Proof:\nopunct] See Appendix \ref{Theorem1_proof}. \end{proof}


The final expression  of $\mathcal{P}_{\mathrm{UL}}$  with fractional pathloss compensation power control is given by
\begin{eqnarray}
\begin{aligned}
\mathcal{P}_{\mathrm{UL}} (\tau) &= \int \limits_0^{\infty} \exp\left( \frac{-\tau \sigma_m^2 l^ {1 - \epsilon}}{\mathrm{P}_{um}  \psi_m} \right) \exp\left( \frac{-2\pi \lambda_m}{\alpha_m} \int \limits_l^{\infty} \left( 1 - \int \limits_0^{\infty}\frac{2\pi \lambda_m u^ {\frac{2}{\alpha_m}-1} e^ {-\pi \lambda_m u^ {\frac{2}{\alpha_m}}}}{\alpha_m (1+\tau l^{1- \epsilon} u^{\epsilon} t^{-1})} \mathrm{d}u \right) t^{\frac{2}{\alpha_m}-1} \mathrm{d}t \right)\\ & \times \bar{F}_s (a_{\mathrm{UL}} l) f_m(l) \mathrm{d}l +  \int \limits_0^{\infty} \exp\left( \frac{-\tau \sigma_s^2 l^{1 - \epsilon}}{\mathrm{P}_{us}  \psi_s} \right)  \bar{F}_m \left(\frac{l}{a_{\mathrm{UL}}}\right) f_s(l) \mathrm{d}l,
\end{aligned}
\end{eqnarray}
where $\epsilon$ is the pathloss compensation factor. The proof for $\mathcal{P}_{\mathrm{UL}, m}$  follows along the same lines as in \cite{NovDhi13} therefore the proof is omitted. The inclusion of the power control adds an extra integral to the Mcell coverage expression which makes it quite complex. Therefore we stick to the assumption of no power control and use the expression in Theorem 1.

\subsection{Rate coverage}
In order to derive the rate coverage, the load on both Mcell and Scell tiers needs to be characterized. We resort to the same approximation used in Section  \ref{Max_rate_sec} where the load is given by (\ref{load_approx}). This approximation is validated with simulation results in Fig. \ref{fig:SINR_rate_cov}(\subref{fig:rate_cov_1}).

%

\textbf{Definition 3.} The rate coverage probability is defined as
\begin{eqnarray*}
\mathcal{R}(\rho)=  \mathbb{P} (R > \rho) =   \mathbb{P} \left(\frac{\mathrm{W}}{N}\log_2(1+ \mathtt{SINR} ) > \rho \right) = \mathbb{P} \left( \mathtt{SINR} > 2^{\frac{\rho N}{\mathrm{W}}} - 1\right).
\end{eqnarray*}

Using the above definition, the UL and DL rate coverage probabilities are given by
\begin{eqnarray}
\mathcal{R}_{\mathrm{DL}} (\rho) = \mathcal{R}_{\mathrm{DL} , m} (\rho) + \mathcal{R}_{\mathrm{DL} , s} (\rho) = \mathcal{P}_{\mathrm{DL} , m} \left(2^{\frac{\rho \bar{N}_{\mathrm{DL} , m}}{\mathrm{W}_m}}-1 \right)+ \mathcal{P}_{\mathrm{DL} , s} \left(2^{\frac{\rho \bar{N}_{\mathrm{DL}, s}}{\mathrm{W}_s}} -1 \right)
\end{eqnarray}
\begin{eqnarray}
\mathcal{R}_{\mathrm{UL}} (\rho) = \mathcal{R}_{\mathrm{UL} , m} (\rho) + \mathcal{R}_{\mathrm{UL} , s} (\rho) = \mathcal{P}_{\mathrm{UL} , m} \left(2^{\frac{\rho \bar{N}_{\mathrm{UL} , m}}{\mathrm{W}_m}}-1 \right)+ \mathcal{P}_{\mathrm{UL} , s} \left(2^{\frac{\rho \bar{N}_{\mathrm{UL} , s}}{\mathrm{W}_s}} -1\right).
\end{eqnarray} 

\section{Performance evaluation and trends}
 \label{validation}
In this section we validate our analysis with Monte Carlo simulations where in each simulation run, UEs and BSs are dropped randomly according to the corresponding densities. All UEs are assumed outdoor. The association criteria, propagation and blockage model are as described in Section \ref{system_model}  and the simulation parameters follow Table \ref{tab:Notation_1}.  Using the analytical results, the different factors that affect the association probability  are studied. Special emphasis is  placed on the Downlink and Uplink Decoupling (DUDe) \cite{ElsBoc14} to understand if decoupling is still useful in the case of mmWave networks. Furthermore, the $\mathtt{SINR}$ and rate coverage trends are illustrated considering the special case of biased DL received power association where the effect of small cell biasing on both $\mathtt{SINR}$ and rate trends is studied with a reflection on the implications in real deployments.

   The parameter values in Table \ref{tab:Notation_1} are used as a baseline. Some of the parameters are altered in some figures in order to understand their effect on the association probability.

\subsection{Association probability}
 \label{Assoc_results1}
We start by looking into the Max-BRP association probabilities derived in Section~\ref{Max_RP} and the different factors that affect these probabilities. It is assumed that $G_s = G_{s_{max}}$ in the association phase. The UL and DL association bias values are unity (0 dB) unless otherwise stated.

\begin{figure*}
	\begin{subfigure}{.5\textwidth}
  \centering
  \includegraphics[width=1.09\linewidth]{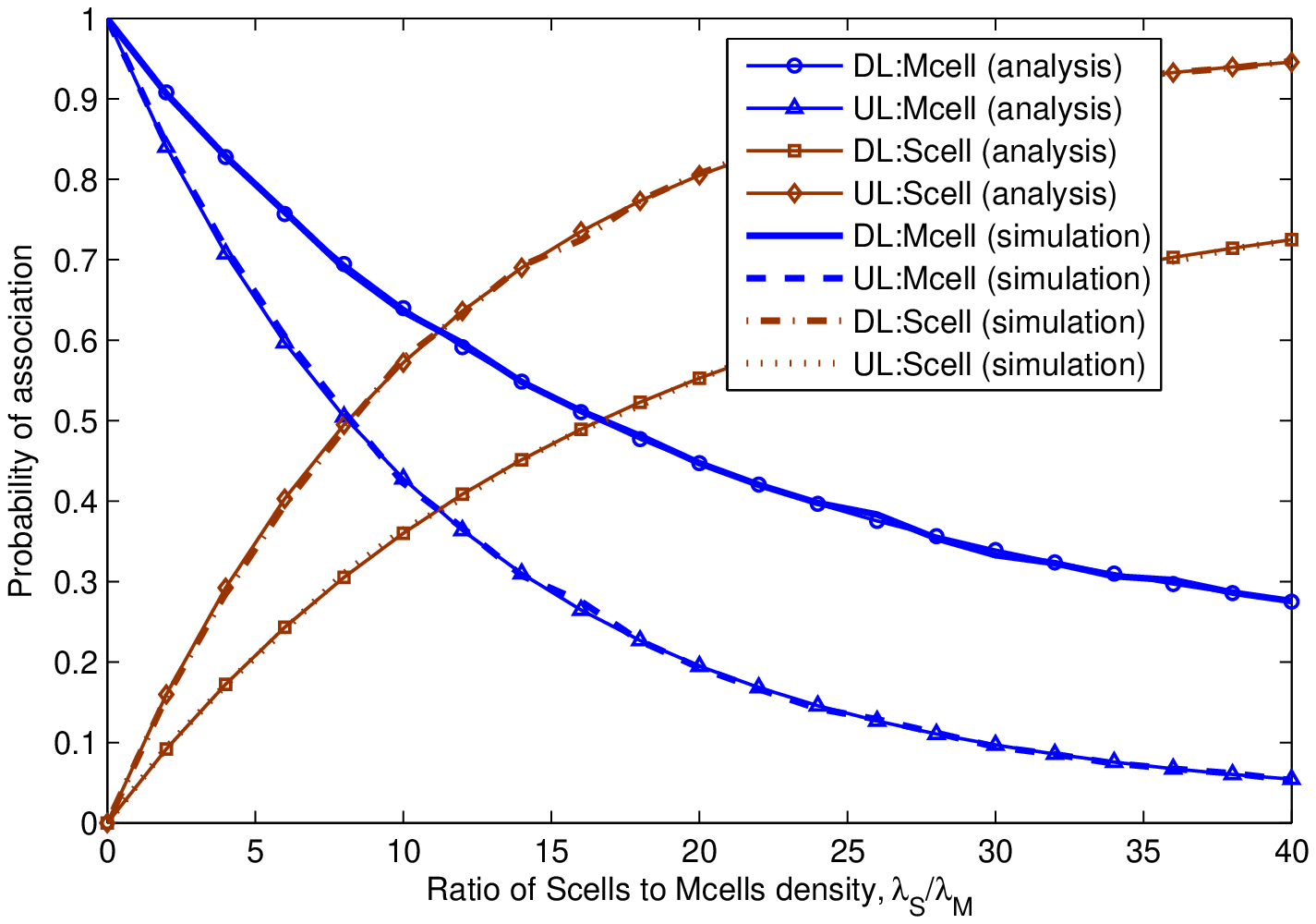}
  \caption{}
  \label{fig:AsspcoProb_2_a}
\end{subfigure}%
\begin{subfigure}{.5\textwidth}
  \centering
  \includegraphics[width=1.09\linewidth]{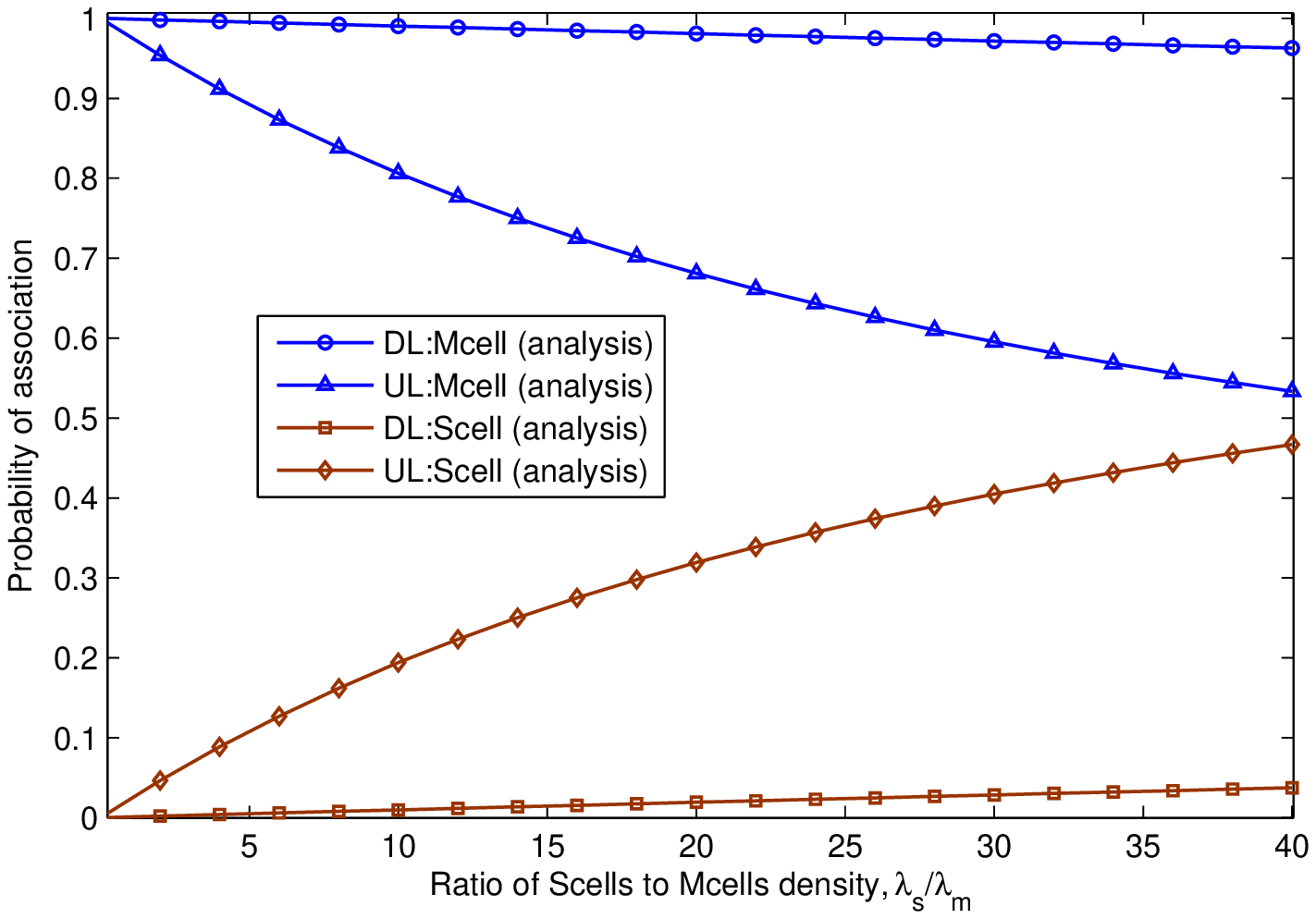}
  \caption{}
  \label{fig:AsspcoProb_2_b}
\end{subfigure}
		\caption{(a) Association probability and validation of the analysis with simulation results for $G_s = 23  \ \mathrm{dBi}$. (b) Association probability with $G_s = 0  \ \mathrm{dBi}$.}
		\label{fig:AssocProb_2_main}
\end{figure*}

\textbf{Association analysis validation.} Fig. \ref{fig:AssocProb_2_main}(\subref{fig:AsspcoProb_2_a}) illustrates the association probabilities derived in Lemma 2 against the ratio of Scells to Mcells density and compared with simulation results. It can be seen that the simulation and analysis results have a very close match which validates our analysis and gives confidence in using the analysis for the following results.
Furthermore, there is a difference between the DL and UL association probabilities for Scells and Mcells, this difference represents the decoupled access where UEs prefer to connect to different cells in the UL and DL. We refer to this difference as the \textit{decoupling gain} for the rest of the paper. In the Scell case, it can be noticed that the UL association probability is always higher than the DL one, this is because the UL coverage of Scells is larger than its DL coverage and vice versa with Mcells. The figure shows that more than 20\% of the UEs have decoupled access at a ratio of Scells to Mcells of 40, in other words the decoupling gain is 20\%.

\textbf{Antenna gain's effect on the association probability}. Fig. \ref{fig:AssocProb_2_main}(\subref {fig:AsspcoProb_2_b}) shows the association probability where $G_s = 0 \ \mathrm{dBi}$, i.e. there is no antenna gain. Predictably, there is very low Scell association probability in the UL and DL. On the other hand, Fig. \ref{fig:AssocProb_2_main}(\subref{fig:AsspcoProb_2_a}) shows a high Scell association probability with $G_s = 23 \ \mathrm{dBi}$.

Another observation is that higher  $G_s$ leads to lower decoupling gain. This stems from the blocking model which is represented by a LOS ball. Most of the DL coverage is inside the LOS ball (with a certain probability of low pathloss exponent (PLE) $(\alpha_l = 2) $) while the UL coverage extends to the NLOS area (with higher PLE $(\alpha_n = 4)$). Therefore increasing the antenna gain expands the DL coverage at a faster rate than the UL coverage due to the difference in the PLE between the LOS and NLOS areas. This in effect reduces the difference between the UL and DL coverage of Scells which, in turn, reduces the decoupling gain. It is worth noting that this trend could be seen with other blockage models as it only depends on the fact that UEs that are closer to the mmWave Scells have higher LOS probability than UEs that are further away which is a general characteristic that would be included in most blockage models.

\begin{figure}
	\centering
		\includegraphics[width=0.6\columnwidth]{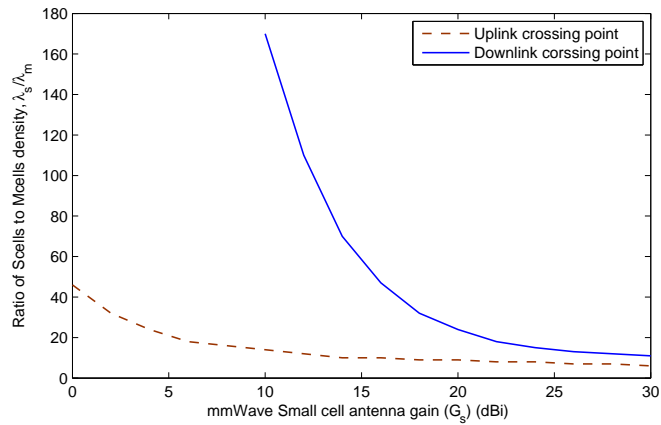}
		\caption{The density $\lambda_s/ \lambda_m$ at which the crossing between the Mcell and Scell UL and DL association probability curves occurs versus the Scell antenna gain $G_s$.}
		\label{fig:crossing_point}
\end{figure}

Fig. \ref{fig:crossing_point} illustrates the ratio of Scell to Mcell density $\lambda_s/\lambda_m$ at which the crossing point between the Scell and Mcell UL and DL association curves occurs versus the antenna gain $G_s$. The difference between the two curves  is an indication of the decoupling gain which is shown to decrease with the increase of $G_s$ which confirms the trend in Fig. \ref{fig:AssocProb_2_main}. In addition, the decreasing tendency of the curves indicate that the crossing point happens at a lower density of Scells the more the gain is increased which results in more and more UEs associating to the Scells.

\begin{figure}
  \centering
  \includegraphics[width=0.6\columnwidth]{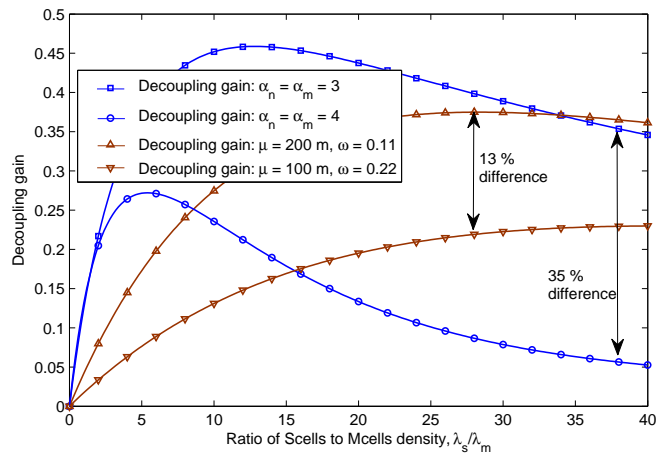}
		\caption{Variation of the decoupling gain with the pathloss exponents ($\alpha$) and the LOS ball parameters.   }
		\label{fig:decoupling_gain}
\end{figure}

\textbf{Pathloss exponent and LOS ball parameters effect on the association probability}. Fig. \ref {fig:decoupling_gain} shows the effect of the PLE and the LOS ball parameters on the decoupling gain. Having a  higher $\alpha_n$ and $\alpha_m$ than $\alpha_l$ results, as in the previous figure, in reducing the difference between the DL and UL coverages of the Scell since the DL coverage is assumed mostly in the LOS ball which makes the DL coverage expand as $\alpha_l$ gets smaller resulting in reducing the gap between the DL and UL coverage borders and, in turn, decreasing the decoupling gain. Therefore the higher the difference between  $\alpha_n$ or $\alpha_m$ with $\alpha_l$, the lower the decoupling gain.

 On the other hand, having a higher LOS ball radius ($\mathrm{\mu}$) results into a higher decoupling gain since as $\mathrm{\mu}$ gets larger more of the UL coverage of Scells area is included in the LOS region which helps in expanding the UL coverage of Scells and, in turn, increases the decoupling gain.
The lower PLE and larger $\mu$ are characteristics of a low density urban environment and indicate that decoupling is more relevant in such a scenario. 

\begin{figure*}
	\begin{subfigure}{.5\textwidth}
  \centering
  \includegraphics[width=1.09\linewidth]{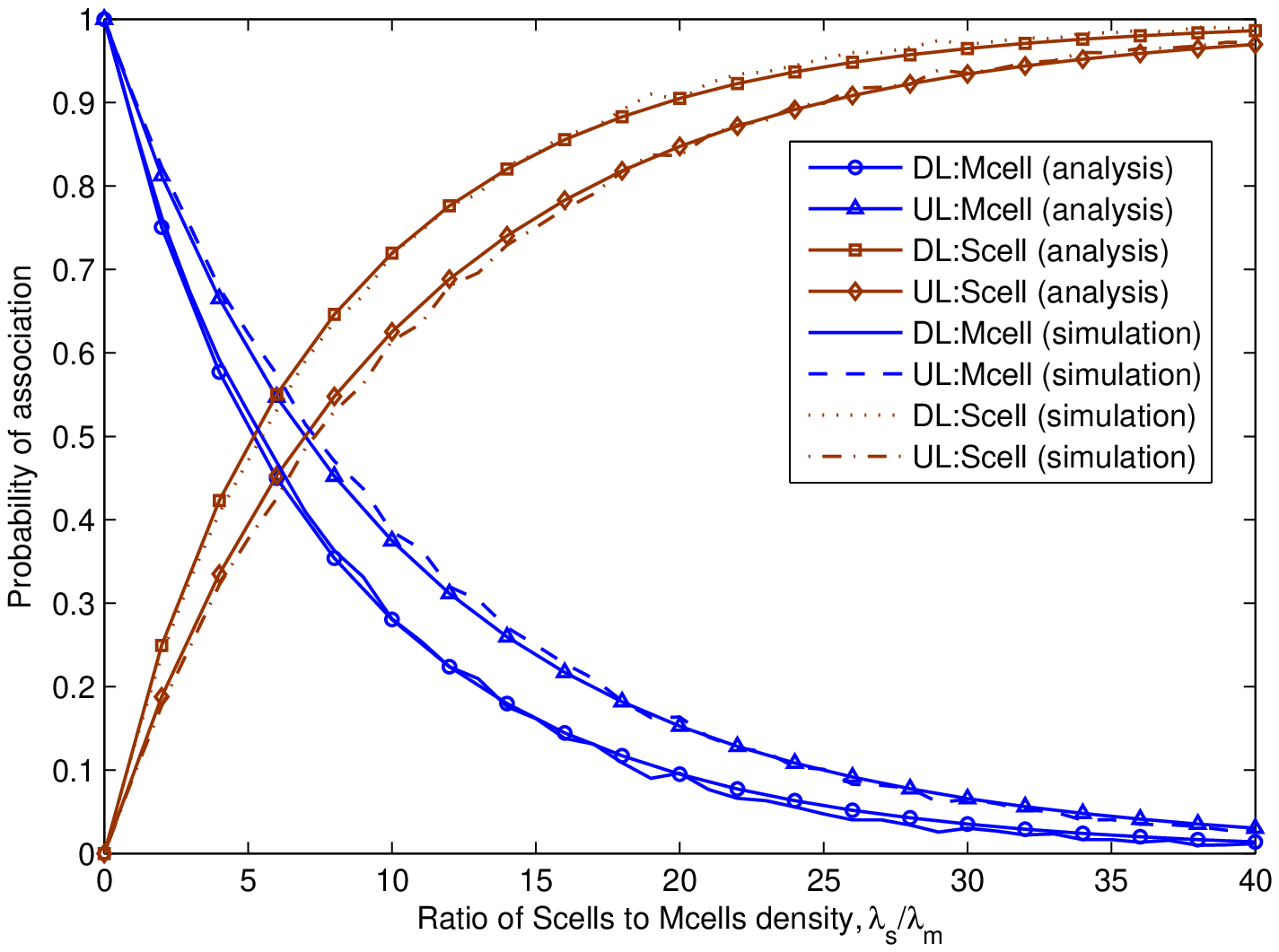}
  \caption{}
  \label{fig:AsspcoProb_r}
\end{subfigure}%
\begin{subfigure}{.5\textwidth}
  \centering
  \includegraphics[width=1.09\linewidth]{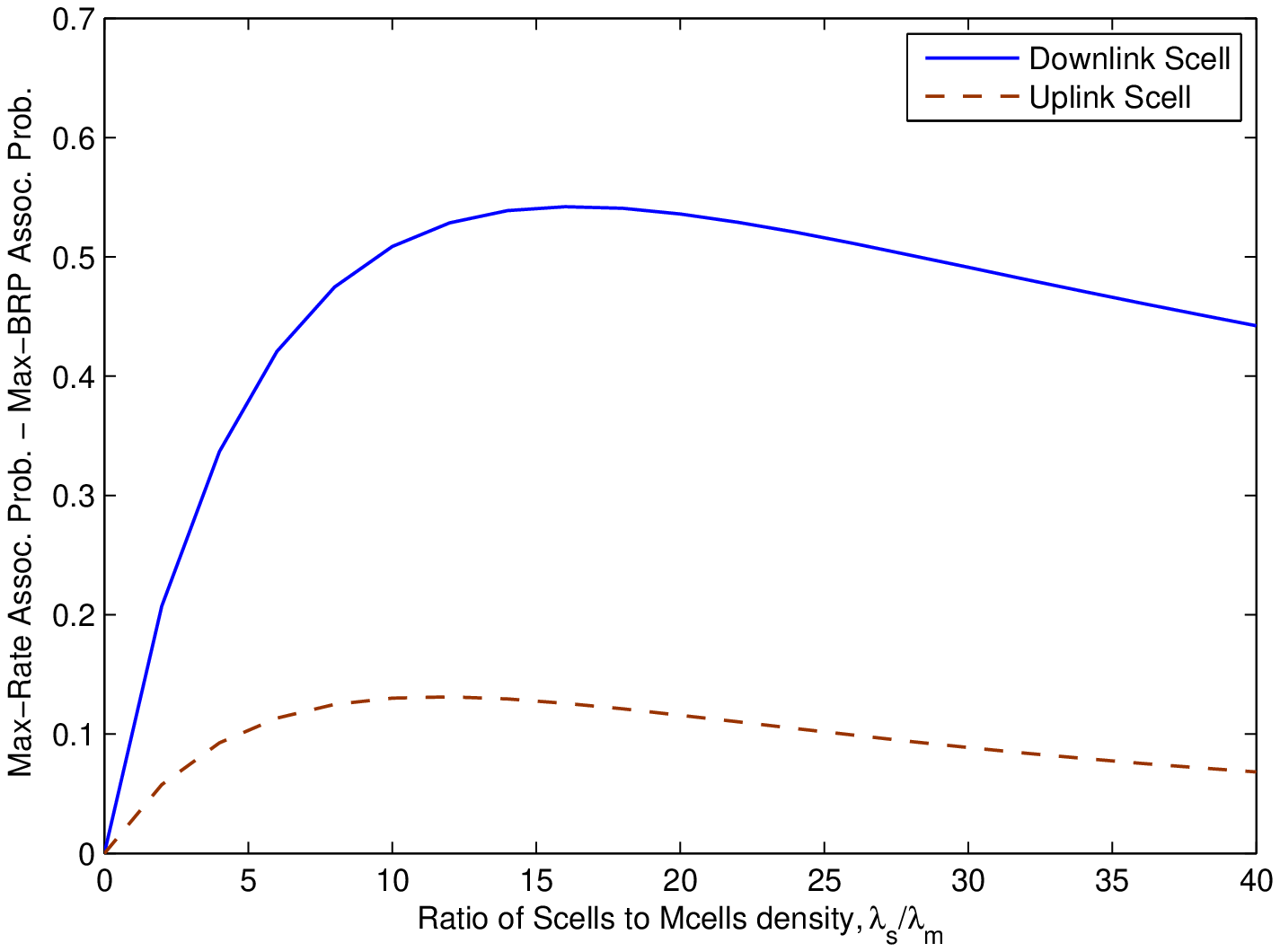}
  \caption{}
  \label{fig:AsspcoProb_r2}
\end{subfigure}
		\caption{(a) Max-Rate Association probability analysis plot compared with simulation. (b) The difference between the UL/DL Scell association probability based on Max-Rate and Max-BRP.  }
		\label{fig:AssocProb_3_main}
\end{figure*}

\textbf{Max-Rate association probability validation and trends}. The results in Fig. \ref{fig:AssocProb_3_main} are based on the Max-Rate association probability derived in Section \ref{Max_rate_sec}. Fig. \ref{fig:AssocProb_3_main}(\subref{fig:AsspcoProb_r}) illustrates the comparison between the analysis and simulation where  the very close match between them validates our analysis and the assumption of having the same exclusion region for UL and DL. The rate based association results into more offloading of UEs towards mmWave Scells as compared to Max-BRP in Fig. \ref{fig:AssocProb_2_main}. This is a direct result from the much wider bandwidth at mmWave Scells. It can also be noticed that there is a decoupling gain in the Max-Rate association as well. However, in this case the decoupled association results from UEs tending to connect to a Scell in the DL and to a Mcell in the UL which is shown by the superior Scell DL association probability compared to the UL and vice versa with the Mcell case. This behaviour is opposite to  the Max-BRP association in Fig. \ref{fig:AssocProb_2_main}. This is a result of the higher bandwidth at mmWave Scells which pushes more UEs to connect to the Scells in the UL and DL  and since --in general-- the UL range is more limited than that of the DL then the mmWave Scells can afford to serve more UEs in the DL than in the UL. This effect is amplified the further the UEs are from the Scell. At a certain point the UL connection towards the Scell is too weak whereas the DL one is relatively stronger  and this is the point where the decoupling happens. 

This effect is further clarified in Fig. \ref{fig:AssocProb_3_main}(\subref{fig:AsspcoProb_r2}) where the increase in the DL association probability in the Max-Rate case over the Max-BRP case is more than 40\% higher than the UL increase.
This important result will be further confirmed in the subsequent results.
\subsection{SINR and rate coverage results}
In this part we present several results for the $\mathtt{SINR}$ and rate coverage to illustrate the effect that the mixed sub-6GHz/mmWave deployment has on the $\mathtt{SINR}$ and rate distributions and how the bias can affect these distributions. From this point onwards, we consider that $\mathrm{T}'_s = \frac{\mathrm{P}_s \mathrm{T}_s}{\mathrm{P}_{us}}$ and  $\mathrm{T}'_m = \frac{\mathrm{P}_m \mathrm{T}_m}{\mathrm{P}_{um}}$ where $\mathrm{T}'_m = \mathrm{T}_{m} = 0  \ \mathrm{dB}$. In other words, we assume that the UL and DL cell associations are based on the DL biased received power where biasing is only assumed for Scells. The reason behind this is to clearly show the effect of small cell biasing on both UL and DL $\mathtt{SINR}$  and rate distributions based on the same association mechanism used in LTE systems where biasing is done jointly for UL and DL and is based on biasing the DL received power. This will help in drawing conclusions related to how currently deployed systems need to be changed and this setup will also be used to confirm our insights regarding the Max-Rate association as will be shown later on.

\begin{figure*}
	\begin{subfigure}{.5\textwidth}
  \centering
  \includegraphics[width=1.09\linewidth]{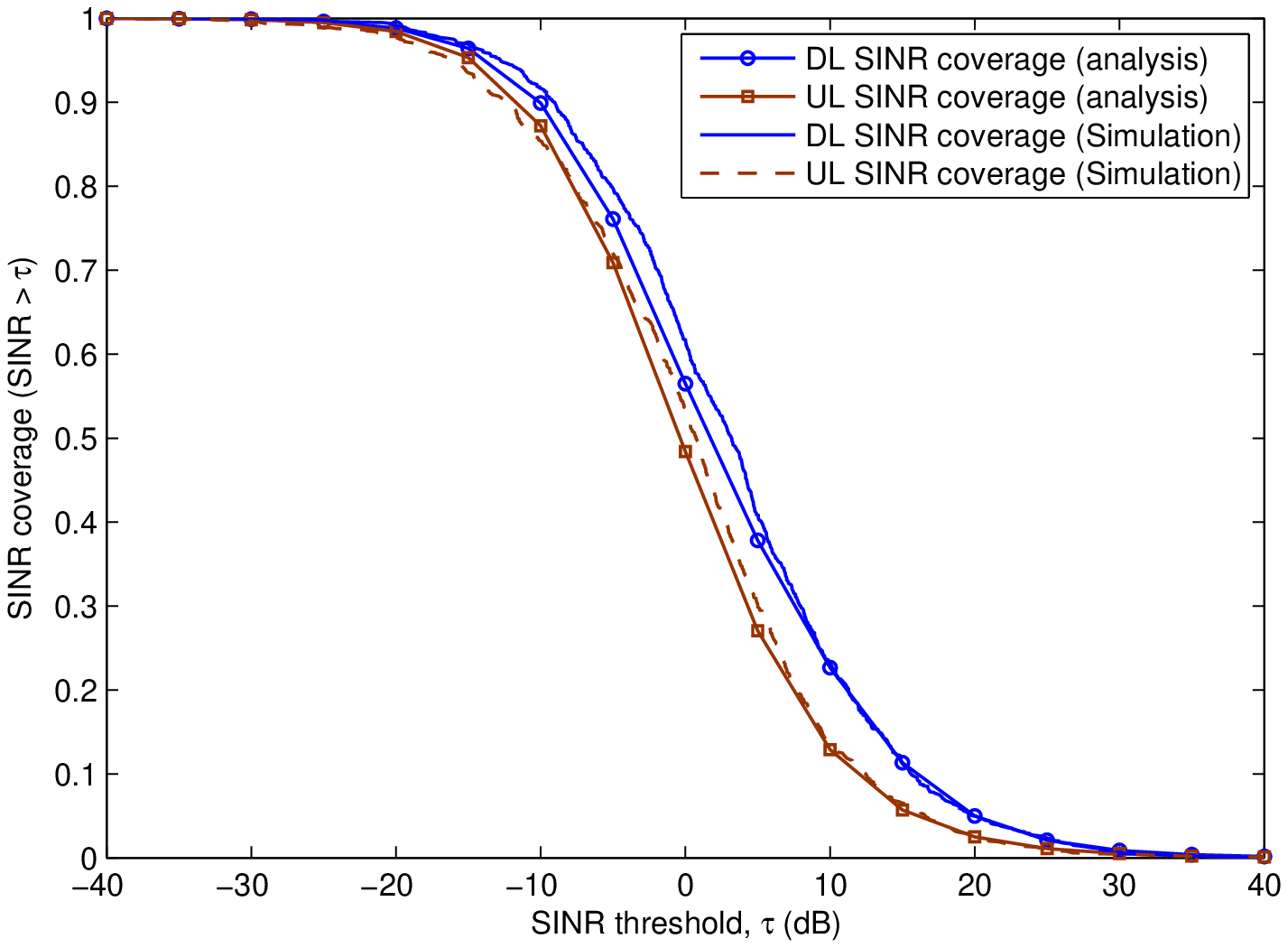}
  \caption{$\mathtt{SINR}$  coverage}
  \label{fig:SINR_cov_1}
\end{subfigure}%
\begin{subfigure}{.5\textwidth}
  \centering
  \includegraphics[width=1.09\linewidth]{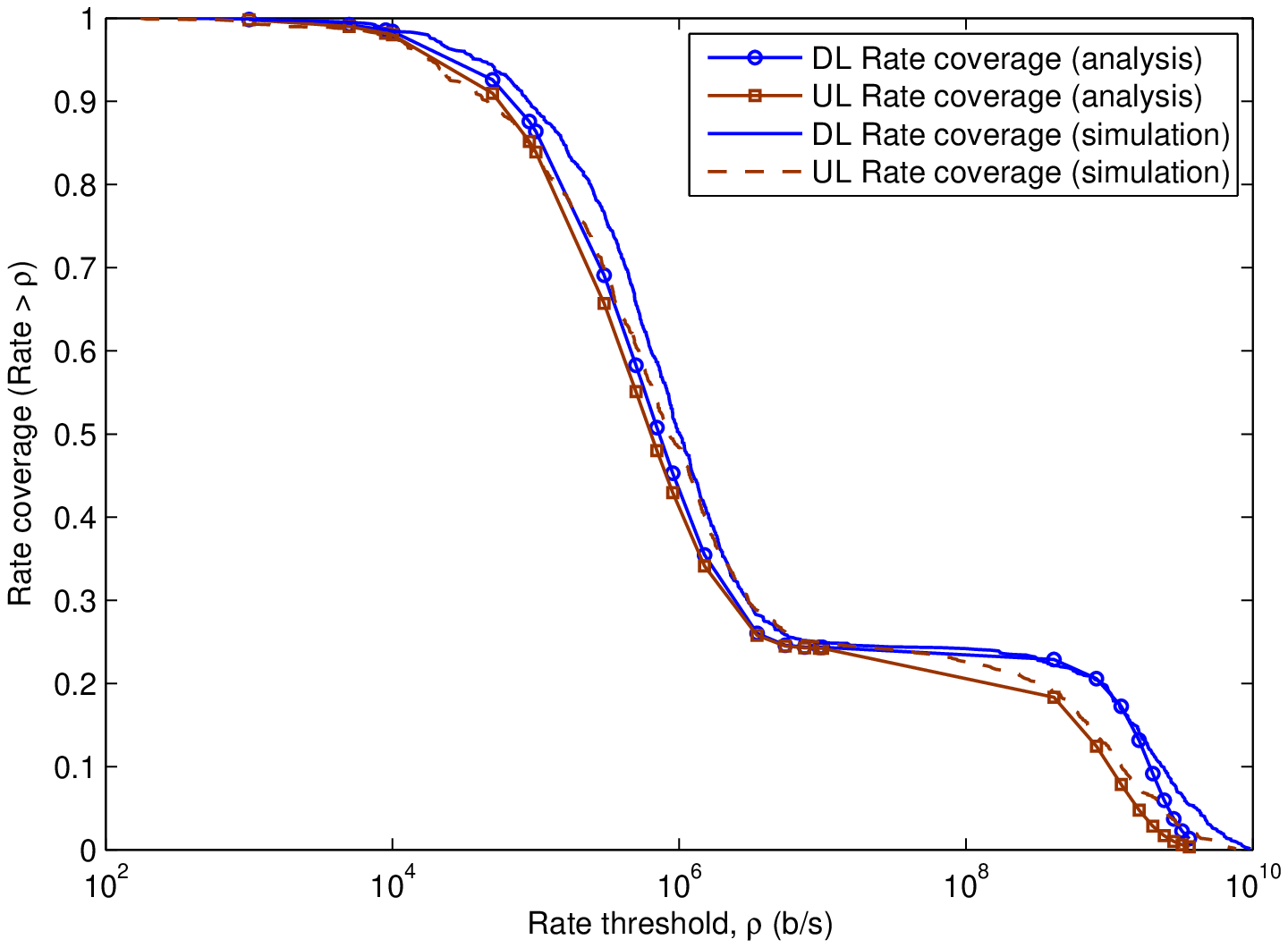}
  \caption{Rate coverage}
  \label{fig:rate_cov_1}
\end{subfigure}
		\caption{$\mathtt{SINR}$ (a) and rate (b) distribution comparison from simulation and analysis.  }
		\label{fig:SINR_rate_cov}
\end{figure*}

\textbf{SINR and rate coverage analysis validation}. Fig. \ref{fig:SINR_rate_cov} shows the $\mathtt{SINR}$ and rate distributions with no bias ($\mathrm{T}_s$ = 0 dB) where the $\mathtt{SINR}$ and rate analysis expressions in Section \ref{SINRratederv} are compared with simulation results. The figure shows that the analysis gives quite accurate results that match very well the simulation results, this allows us to use the analysis for further insights in the coming results. Furthermore, Fig. \ref{fig:SINR_rate_cov}(\subref{fig:rate_cov_1}) has a flat area between $10^7$ and $10^9$(b/s) rate threshold, this area separates the sub-6GHz UEs  below $10^7$(b/s) from the mmWave UEs with very good channel above $10^9$(b/s). This shows the substantial difference in rate that the larger bandwidth in mmWave could offer.

\begin{figure}
	\centering
		\includegraphics[width=0.6\columnwidth]{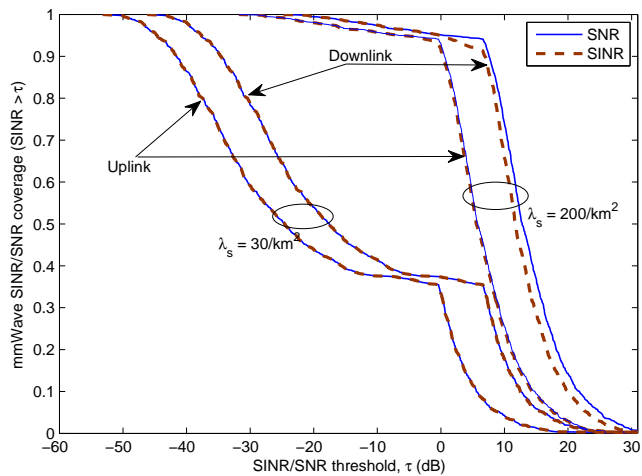}
		\caption{Simulation results for the distribution of the mmWave $\mathtt{SINR}$ and $\mathtt{SNR}$ which validates the assumption ($\mathtt{SINR}$ $\approx$ $\mathtt{SNR}$). }
		\label{fig:SINR_SNR}
\end{figure}

\textbf{(SINR $\approx$ SNR) validation}. Fig. \ref{fig:SINR_SNR} shows simulation results for the CCDF of the mmWave UEs UL and DL $\mathtt{SINR}$ and $\mathtt{SNR}$ for two different mmWave Scells densities. It can be seen from the figure that for $\lambda_s = 30/\mathrm{km^2}$ the $\mathtt{SINR}$ and $\mathtt{SNR}$ are almost overlapping and even at $\lambda_s = 200/\mathrm{km^2}$ the difference between $\mathtt{SINR}$ and $\mathtt{SNR}$ is very small. This result confirms our assumption that interference in mmWave has a minimal impact on coverage for the mmWave Scells densities considered in our scenario. This, in turn, confirms that the $\mathtt{SINR}$ can be approximated by the $\mathtt{SNR}$ for mmWaves which is quite different than the trend in sub-6GHz networks where $\mathtt{SINR} \approx \mathtt{SIR}$. Furthermore, the break point in the curves at 30\% and 90\% of the CCDF for $\lambda_s$ of 30 and 200 shows how the $\mathtt{SNR}$ starts degrading quickly after a certain point which is a result of the LOS ball blockage model which assumes that beyond a certain distance between the UE and the BSs all the UEs are considered non line of sight. In addition, the degradation affects fewer UEs at $\lambda_s = 200/\mathrm{km^2}$ since at a higher density fewer UEs are expected to be outside the LOS ball of the mmWave Scells.

\textbf{Scell biasing effect on SINR and rate trends}. Several previous studies have shown the importance of cell biasing in Hetnets \cite{SinAnd13, YeChe13, WanPed12}. However, the different propagation characteristics of sub-6GHz and mmWaves and the high imbalance in the available resources in both bands could result in different conclusions when it comes to biasing. Hence, the following results focus on the effect of biasing on the system's $\mathtt{SINR}$ and rate coverage and the optimal value of biasing for UL and DL.

\begin{figure*}
	\begin{subfigure}{.5\textwidth}
  \centering
  \includegraphics[width=1.09\linewidth]{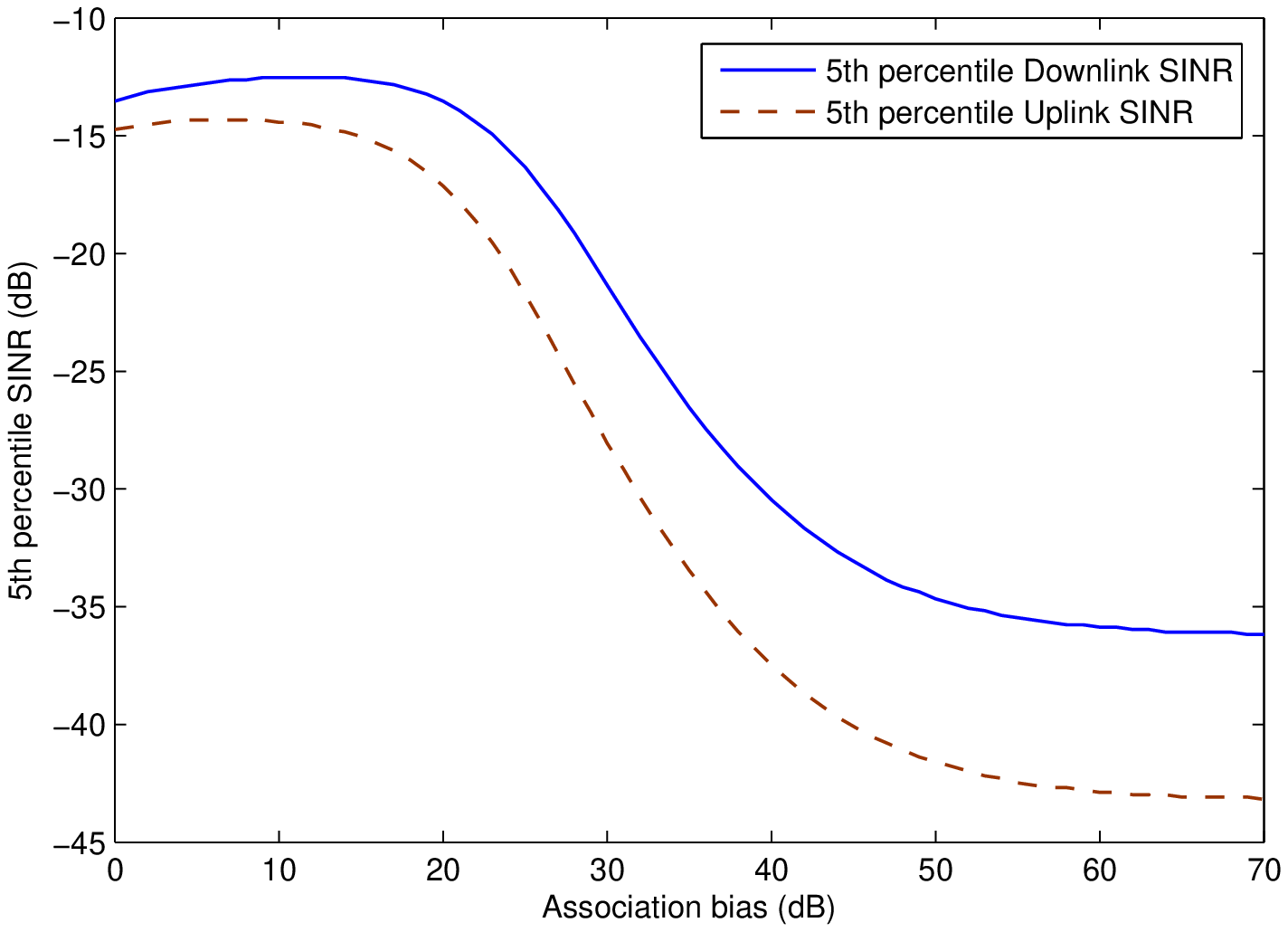}
  \caption{5th percentile $\mathtt{SINR}$}
  \label{fig:5th_SINR}
\end{subfigure}%
\begin{subfigure}{.5\textwidth}
  \centering
  \includegraphics[width=1.09\linewidth]{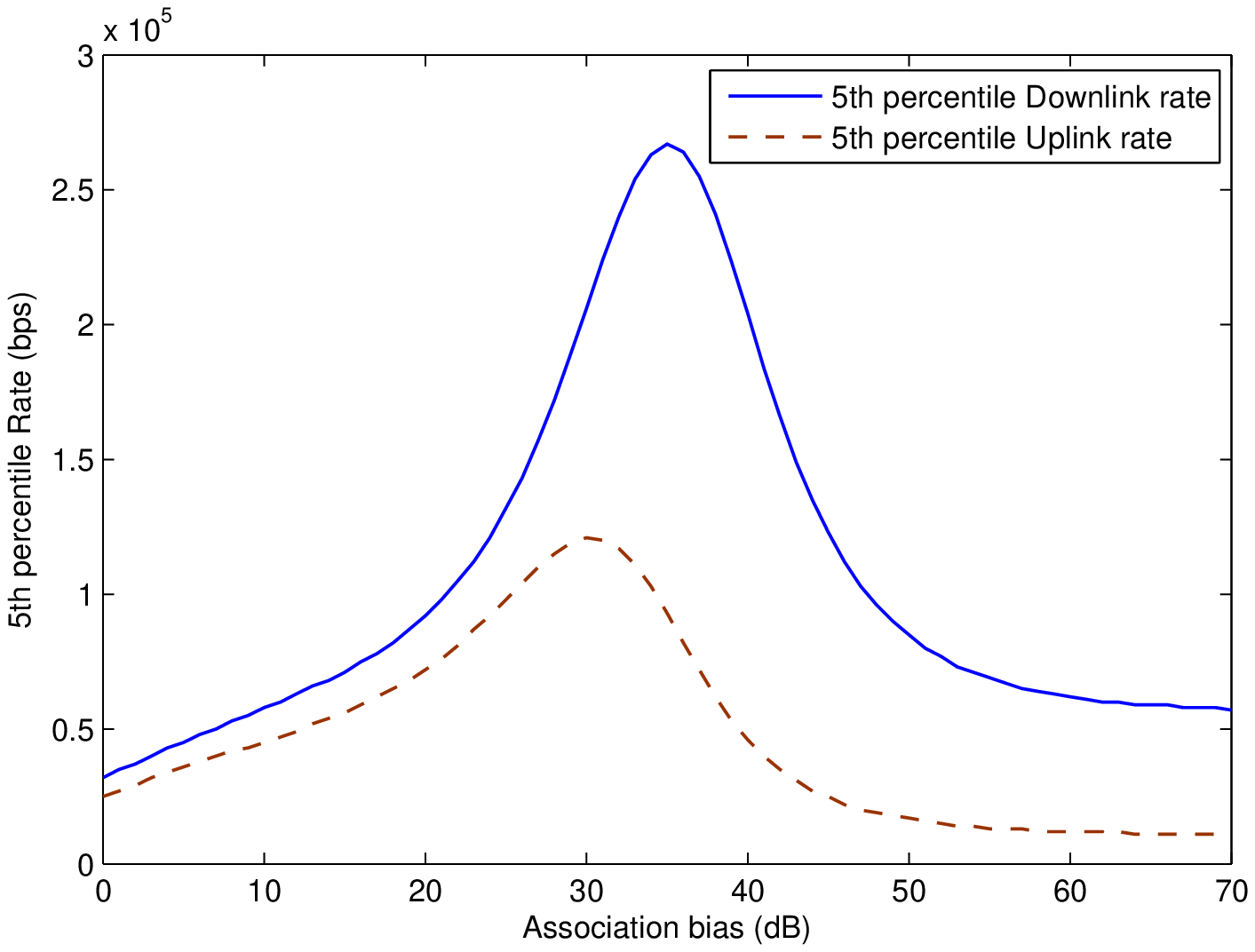}
  \caption{5th percentile rate}
  \label{fig:5th_rate}
\end{subfigure}
		\caption{5th percentile $\mathtt{SINR}$ (a) and rate (b) against the small cell bias value in dB. }
		\label{fig:SINR_rate_sim1}
\end{figure*}

 Fig. \ref {fig:SINR_rate_sim1} illustrates the UL and DL 5th percentile $\mathtt{SINR}$ $\tau_{95}$ and rate $\rho_{95}$ against the Scell association bias ($\mathrm{T}_s$) where the relation between the 5th percentile $\mathtt{SINR}$ and the $\mathtt{SINR}$ coverage is ($\mathcal{P}(\tau_{95}) = 0.95$) and the same for rate. In Fig. \ref {fig:SINR_rate_sim1}(a) the DL $\mathtt{SINR}$ increases slightly and then starts decreasing beyond 5 dB bias, the UL $\mathtt{SINR}$ behaves similarly. The slight increase in $\mathtt{SINR}$ at the beginning is due to the negligible interference in mmWave networks, therefore although the $\mathtt{SNR}$ is reduced the overall $\mathtt{SINR}$ is slightly increased.
On the other hand, the 5th percentile rate in Fig.~\ref {fig:SINR_rate_sim1}(b) is peaking at a bias of 30 and 35 dB for the UL and DL respectively. However,  the corresponding $\mathtt{SINR}$ values with such large bias are around -30 dB, which is extremely low. The reason for the high rate despite the low $\mathtt{SINR}$ is obviously the much higher bandwidth at mmWaves. These bias values are over 100x the typical values seen in sub-6GHz scenarios in \cite{SinAnd13, YeChe13, WanPed12}.

The design insight behind this result is that very robust modulation and coding schemes need to be considered for mmWave networks so that they can operate at very low $\mathtt{SINR}$. Another insight is that the UL 5th percentile rate peaks at a lower bias value than the DL rate which means that a fraction of the UEs would tend to connect to the Scell and Mcell in the DL and UL respectively. This confirms the trend resulting from the Max-Rate association in Fig. \ref{fig:AssocProb_3_main}(\subref{fig:AsspcoProb_r}) about the  \textit{reversed} decoupling behaviour since at the optimal bias value UEs are assumed to be connected to their rate optimal cell. This also confirms that the association probability in Section \ref{Max_rate_sec} (with $N_m = N_s = 1$) results in the same trend as the optimal rate results (considering the cell loads) in Fig. \ref{fig:SINR_rate_sim1}(b).

\begin{figure*}
	\begin{subfigure}{.5\textwidth}
  \centering
  \includegraphics[width=1.09\linewidth]{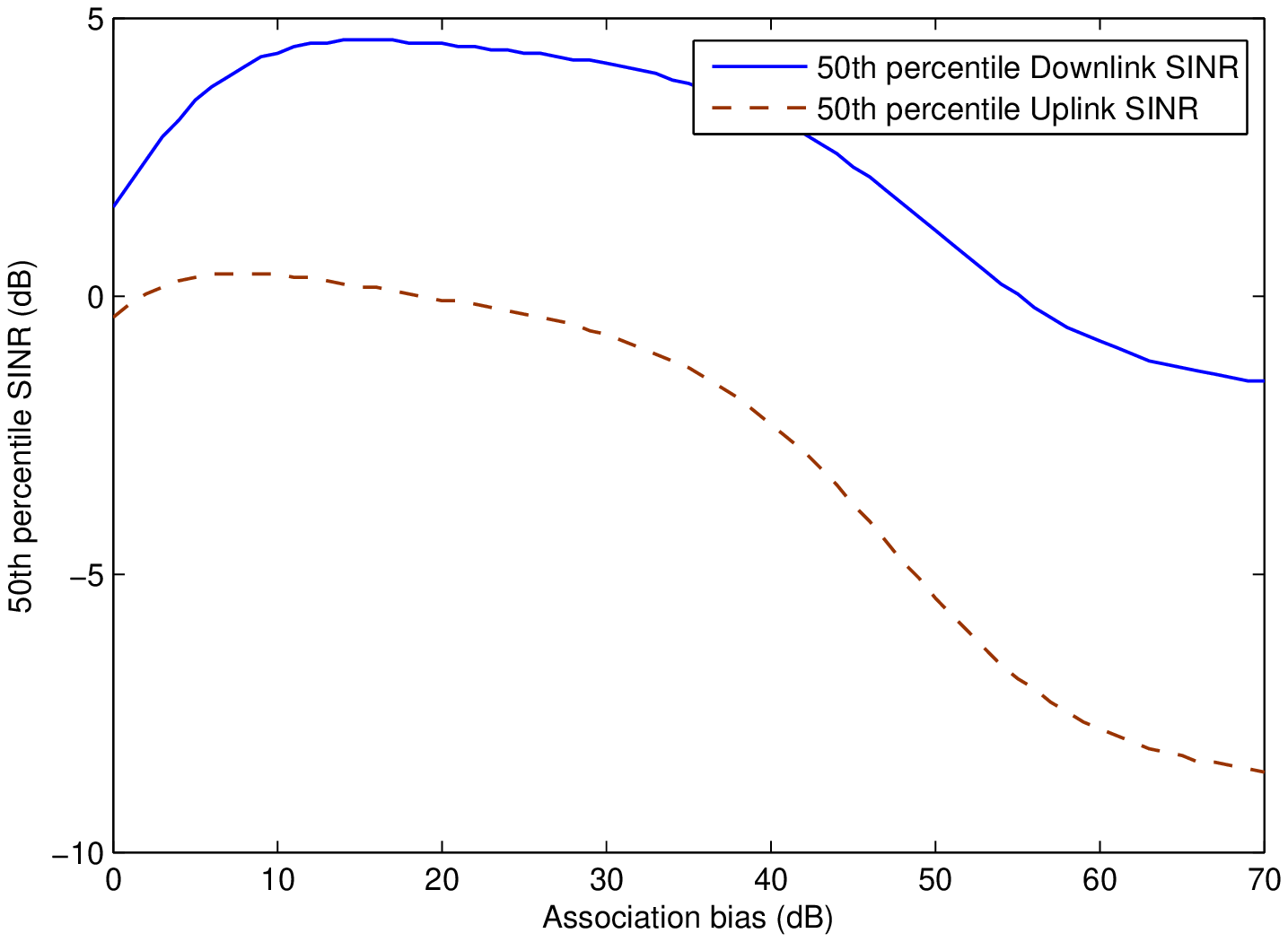}
  \caption{50th percentile $\mathtt{SINR}$}
  \label{fig:50th_SINR}
\end{subfigure}%
\begin{subfigure}{.5\textwidth}
  \centering
  \includegraphics[width=1.09\linewidth]{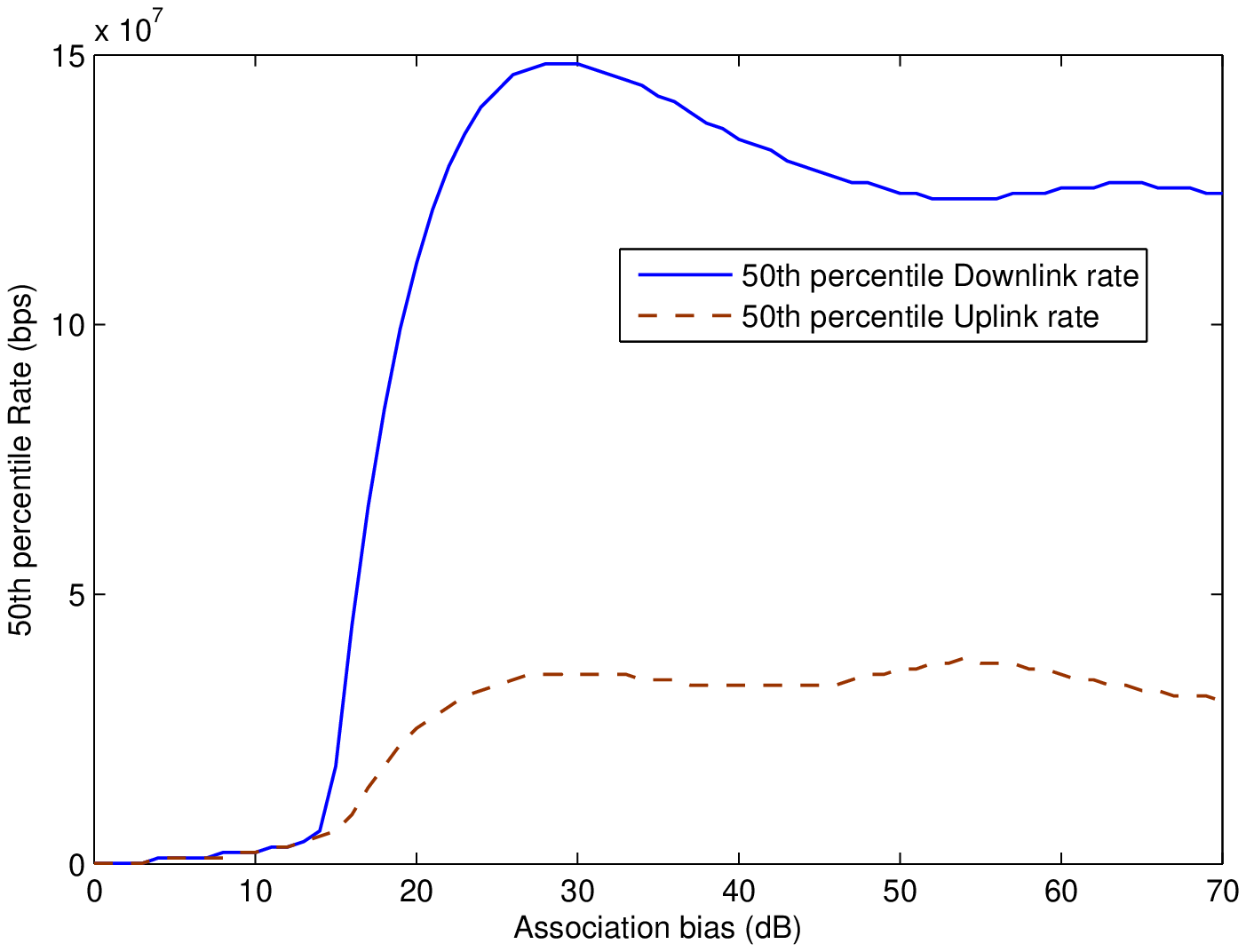}
  \caption{50th percentile rate}
  \label{fig:50th_rate}
\end{subfigure}
		\caption{50th percentile $\mathtt{SINR}$ (a) and rate (b) against the small cell bias value in dB. }
		\label{fig:SINR_rate_sim2}
\end{figure*}

Fig. \ref {fig:SINR_rate_sim2} illustrates the 50th percentile $\mathtt{SINR}$ and rate where a similar behaviour to  Fig. \ref{fig:SINR_rate_sim1} can be noticed. However the increase in the 50th percentile $\mathtt{SINR}$ in Fig. \ref {fig:SINR_rate_sim2}(a) is much higher than in the 5th percentile $\mathtt{SINR}$, also the decline starts at a higher bias than the 5th percentile $\mathtt{SINR}$, this is because the 50th percentile UEs typically are closer or have a better channel to their serving cells. Therefore, a degradation in their $\mathtt{SINR}$ would require a higher bias value. Looking at the rate in Fig. \ref {fig:SINR_rate_sim2}(b), it can be noticed that it peaks at around 30 dB bias for the UL and DL which corresponds to an $\mathtt{SINR}$ of 3 dB for the DL and -2 dB for the UL which is still considered low for the median UEs, therefore the need for robust modulation and coding still applies in the 50th percentile UEs case. The fluctuation in the 50th percentile rate beyond 50 dB bias results from UEs moving from their serving mmWave Scells to  less loaded mmWave Scells which results in a slight increase in the rate.

\begin{figure}
	\centering
		\includegraphics[width=0.6\columnwidth]{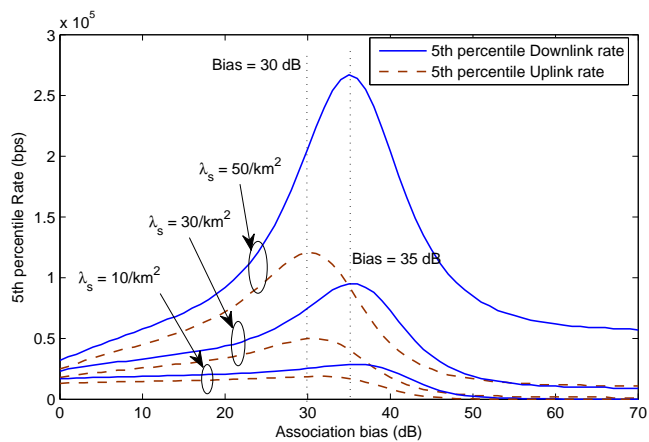}
		\caption{UL and DL 5th percentile rate with variable mmWave Scell densities.}
		\label{fig:5th_percent_rate_density}
\end{figure}

\textbf{Impact on infrastructure density}. In Fig. \ref {fig:5th_percent_rate_density} the impact of the density of mmWave Scells on the 5th percentile UL and DL rate is illustrated. It can be observed that the optimal bias in terms of achieved rate is  30 and 35 dB for the UL and DL respectively and these values are the same for all densities. It has been shown in \cite{SinAnd13} that the optimal bias considering resource partitioning decreases with the increase in the Scells density because of the increased interference on the range expanded UEs. However, in our scenario it was already shown that mmWave operation is noise limited, therefore interference has a marginal effect which explains the invariance of the optimal bias with the Scell density.

\subsection{System design implications}
\label{design_implications}
In this part we summarize the system design and deployment implications based on the results shown previously in this section:
\begin{itemize}
\item The association probability of mmWave Scells is dramatically improved as the Scell beamforming gain is taken into account during the association phase. This indicates the importance of having highly directional beams in the association phase.
\item Decoupled access is still relevant in mmWave/sub-6GHz HetNets from a maximum received power as well as rate based associations. It was shown that DUDe is key for optimal performance in both UL and DL whether the optimization criterion is received power or rate where decoupling occurs in different directions for the two criteria.
\item Decoupled access is more relevant in less dense urban environments. This is reflected in Fig. \ref{fig:decoupling_gain} by the higher decoupling gain with a smaller $\alpha_m$ and $\alpha_n$ and a higher LOS ball radius ($\mathrm{\mu}$) and both features characterise low density urban scenarios.
\item Aggressive values of mmWave Scell biasing can be beneficial in terms of rate as shown in Fig. \ref{fig:SINR_rate_sim1}(b). This would result in UEs having to operate in very low $\mathtt{SINR}$ which gives rise to  a need for robust modulation and coding techniques that would allow the UEs to operate in these low $\mathtt{SINR}$ regimes to harvest the benefits of mmWaves.
\item It was shown that from a rate perspective UEs are more probable to connect to sub-6GHz Mcell in the UL and mmWave Scell in the DL. In addition, recent studies on electromagnetic field exposure \cite{ColThor15} have shown that the maximum UL transmit power on frequencies above 6GHz will need to be several dB lower than sub-6GHz to be compliant with exposure limits. These trends could lead to allocating the UL on sub-6GHz Mcells and the DL on mmWave Scells as discussed in \cite{BocAnd15}.
\end{itemize}

\section{Conclusions}

In this paper we proposed a detailed analytical framework for cell association in sub-6GHz-mmWave heterogeneous networks considering a decoupled uplink and downlink association. The analysis considered a maximum biased received power as well as a maximum achievable rate approach highlighting the main differences between them. The results show that there is a different trend in decoupling between the two approaches where in the rate based approach devices tend to connect in the UL to the sub-6GHz Mcells which is opposite to the decoupling trend in previous studies. The $\mathtt{SINR}$ and rate coverages are also derived where we put special emphasis on Scell biasing in the results showing that quite high Scell bias values are possible which has implications on the modulation and coding schemes in future networks. The presented work could be extended in numerous ways including the consideration of UL power control, indoor users and mobility in a mmWave scenario. Considering the cell load in the rate based association is an interesting extension as well. In addition, the inclusion of sub-6GHz small cells and allowing users to have multiple decoupled connections in the uplink and downlink to different base stations is quite interesting and will be left for future work.

\appendices
\section{}
\begin{proof}[\unskip\nopunct]
\label{lemma1_proof}
\textit{Derivation of Lemma 1:}
Starting with the mmWave case, the propagation process $\mathcal{N}_s :=\{ L_s(x) = \|x\|^{\alpha_s(x)}  \}$ on $\mathbb{R^+}$ for $x \in \Phi_s $ has intensity
\begin{eqnarray*}
\Lambda_s((0,t)] = \int\limits_{\mathbb{R}^2}  \mathbb{P}(L_s(x) < t) \mathrm{d}x = 2\pi\lambda_s \int\limits_0^{\infty} \mathbb{P}(r^{\alpha_s(r)} <t)r \mathrm{d}r.
\end{eqnarray*}
In the previous equation $\alpha$ is distance dependent as it has different values for LOS and NLOS links and according to the blockage model in Section \ref{blockage_model} the intensity can be expressed as 
\begin{eqnarray*}
\Lambda_s((0,t)]&=& 2\pi\lambda_s \left( \mathrm{\omega} \int\limits_0^\mathrm{\mu} r \mathds{1}(r^{\alpha_l} < t) \mathrm{d}r +  (1-\mathrm{\omega}) \int\limits_0^\mathrm{\mu} r \mathds{1}(r^{\alpha_n} < t)\mathrm{d}r + \int\limits_\mathrm{\mu}^{\infty} r \mathds{1}(r^{\alpha_n} < t)\mathrm{d}r \right)\\
&=&  2\pi\lambda_s \left( \mathrm{\omega} \int\limits_0^\mathrm{\mu} r \mathds{1}(r < t^{\frac{1}{\alpha_l}}) \mathrm{d}r +  (1-\mathrm{\omega}) \int\limits_0^\mathrm{\mu} r \mathds{1}(r < t^{\frac{1}{\alpha_n}})\mathrm{d}r + \int\limits_\mathrm{\mu}^{\infty} r \mathds{1}(r < t^{\frac{1}{\alpha_n}})\mathrm{d}r \right)\\
&=&  2\pi\lambda_s \left( \mathrm{\omega} \int\limits_0^{\min(\mathrm{\mu},t^{\frac{1}{\alpha_l}})} r \mathrm{d}r + (1 - \mathrm{\omega}) \int\limits_0^{\min(\mathrm{\mu},t^{\frac{1}{\alpha_n}})} r \mathrm{d}r + \int\limits_\mathrm{\mu}^{t^{\frac{1}{\alpha_n}}} r \mathds{1}(t^{\frac{1}{\alpha_n}} > \mathrm{\mu})\mathrm{d}r \right).
\end{eqnarray*}
Solving the integrals yields
\begin{align*}
\Lambda_s((0,t)] &=   \pi\lambda_s \Bigg( \mathrm{\omega} \left( \mathrm{\mu}^2 \mathds{1} (t>\mathrm{\mu}^{\alpha_l}) + t^{\frac{2}{\alpha_l}} \mathds{1} (t\leq \mathrm{\mu}^{\alpha_l})\right) + (1-\mathrm{\omega})\Big( \mathrm{\mu}^2 \mathds{1} (t>\mathrm{\mu}^{\alpha_n})\\ & + t^{\frac{2}{\alpha_l}} \mathds{1} (t\leq \mathrm{\mu}^{\alpha_n})\Big)
 + (t^{\frac{2}{\alpha_n}} - \mathrm{\mu}^2) \mathds{1} (t>\mathrm{\mu}^{\alpha_n})\Bigg).
\end{align*}
Finally, rearranging the terms yields the final expression for the pathloss process intensity in (\ref{lambda_s}).
For the sub-6GHz Mcells case, deriving the pathloss process intensity is straight forward as blockage is not considered for sub-6GHz.
The propagation process $\mathcal{N}_m :=\{ L_m(x) = \|x\|^{\alpha_m}  \}$ on $\mathbb{R^+}$ for $x \in \Phi_m $ has intensity
\begin{eqnarray*}
\Lambda_m((0,t)] = \int\limits_0^{\infty}\mathbb{P}(L_m(x) < t) \mathrm{d}x = 2\pi\lambda_m \int\limits_0^{\infty} \mathbb{P}(r^{\alpha_m} <t)r \mathrm{d}r
= 2\pi\lambda_m \int\limits_0^{t^{\frac{1}{\alpha_m}}} r \mathds{1}(r < t^{\frac{1}{\alpha_m}}) \mathrm{d}r = \pi \lambda_m t^{\frac{2}{\alpha_m}}.
\end{eqnarray*}
\end{proof}

\section{}
\label{lemma3_proof}
\begin{proof}[\unskip\nopunct]
\textit{Derivation of Lemma 3:}
The derivation of the Max-Rate association probabilities starts with the downlink association probability to a sub-6GHz Mcell $\mathcal{B}_{DL, m}$ which is given by
\begin{align}
\mathcal{B}_{\mathrm{DL}, m} & = \mathbb{P} \left(  \mathtt{SIR}_{\mathrm{DL},m} > (1+\mathtt{SNR}_{\mathrm{DL},s})^{\left(\frac{\mathrm{W}_s}{\mathrm{W}_m}  \right)} -1 \right)\nonumber\\
&= \mathbb{E}_{(\mathtt{SNR}_{\mathrm{DL},s} = S)} \left[\bar{F}_{\mathtt{SIR}_{\mathrm{DL},m}}\left( (1 + S)^{\frac{\mathrm{W}_s}{\mathrm{W}_m}} -1 \right)   \right]
= \int\limits_0^{\infty} f_{\mathtt{SNR}_{\mathrm{DL},s}}(z) \bar{F}_{\mathtt{SIR}_{\mathrm{DL}, m}}\left( (1 + z )^{\frac{\mathrm{W}_s}{\mathrm{W}_m}} -1 \right)\mathrm{d}z\nonumber,
\end{align}
where $\bar{F}_{\mathtt{SIR}_{\mathrm{DL},m}}(k)$ is the DL coverage probability $\mathcal{P}(\mathtt{SIR} > k)$ and $f_{\mathtt{SNR}_{\mathrm{DL},s}}(z)$ is the PDF of $\mathtt{SNR}_{\mathrm{DL},s}$. For $\bar{F}_{\mathtt{SIR}_{\mathrm{DL},m}}(k)$ we use the expression derived in \cite{AndBac11} for the coverage probability in the no noise and exponential fading case which is given by
\begin{eqnarray}
\bar{F}_{\mathtt{SIR}_{\mathrm{DL},m}}(t) = \frac{1}{1+ \rho(t, \alpha_m)},
\end{eqnarray}
where $\rho(t, \alpha_m)= t^{\frac{2}{\alpha_m}} \int\limits_{t^{\frac{-2}{\alpha_m}}}^{\infty} \frac{\mathrm{d}u}{1 + u^{\frac{\alpha_m}{2}}}  $.
$f_{\mathtt{SNR}_{\mathrm{DL},s}}(z)$ can be derived from $\bar{F}_{\mathtt{SNR}_{\mathrm{DL},s}}(z)$ as follows
\begin{eqnarray*}
\bar{F}_{\mathtt{SNR}_{\mathrm{DL},s}}(z) = \mathbb{P} (\mathtt{SNR}_{\mathrm{DL},s} > z) = \mathbb{P} \left(\frac{\mathrm{P}_s \psi_s h_{x^*,0}  L_s(x^*)^{-1}}{\sigma_{s}^2}  > z \right) = \int\limits_0^{\infty} \exp\left(\frac{-z \sigma_s^2 l}{\mathrm{P}_s \psi_s}\right) f_s(l) \mathrm{d}l.
\end{eqnarray*}
\begin{eqnarray*}
f_{\mathtt{SNR}_{\mathrm{DL},s}}(z) = \frac{- \mathrm{d} \ \bar{F}_{\mathtt{SNR}_{\mathrm{DL},s}}(z)}{\mathrm{d} z} = \frac{- \mathrm{d}}{\mathrm{d}z}  \int\limits_0^{\infty} \exp\left(\frac{-z \sigma_s^2 l}{\mathrm{P}_s \psi_s}\right) f_s(l) \mathrm{d}l = \frac{- \mathrm{d} }{dz} \int\limits_0^{\infty} f_s(z, l) \mathrm{d}l,
\end{eqnarray*}
where  $f_s(l)$ is given in (\ref{f_s}). 
In order to simplify the previous expression we exchange the order of the differentiation and integral using Leibnitz Rule \cite{Imai06}. The following two conditions need to be satisfied in order for this rule to be applicable.
\begin{itemize}
\item $ \abs{\frac{\mathrm{d} f_s(z, l)}{\mathrm{d}z}} \leq g(z, l)$, which means that the LHS expression is differentiable.
\item $\int\limits_0^\infty g(z, l) \mathrm{d}z < \infty $, $g(z, l)$ is defined below.
\end{itemize}
\begin{eqnarray*}
 \abs{\frac{\mathrm{d}f_s(z, l)}{\mathrm{d}z}} \leq \frac{\sigma_s^2}{\mathrm{P}_s \psi_s} \ l \ f_s(l) = g(z, l),
\end{eqnarray*}
which satisfies the first condition.
\begin{eqnarray*}
\int\limits_0^\infty g(z, l) \mathrm{d}z &=& \frac{\sigma_s^2}{\mathrm{P}_s \psi_s} \int\limits_0^\infty l f_s(l) \mathrm{d}l = \frac{\sigma_s^2}{\mathrm{P}_s \psi_s} \int\limits_0^\infty \mathbb{P}(L_s > x) \mathrm{d}x = \frac{\sigma_s^2}{\mathrm{P}_s \psi_s} \int\limits_0^\infty e^{-\Lambda_s(0, x)}\mathrm{d}x \\
&\stackrel{(a)}{=}& \int\limits_0^{\mu^{\alpha_l}} e^{-\Lambda_s(0, x)}\mathrm{d}x + \int\limits_{\mu^{\alpha_l}}^{\mu^{\alpha_n}} e^{-\Lambda_s(0, x)}\mathrm{d}x + \int\limits_{\mu^{\alpha_n}}^\infty e^{-\Lambda_s(0, x)}\mathrm{d}x = \mathrm{constant}  + \int\limits_{\mu^{\alpha_n}}^\infty e^{-x^{\frac{2}{\alpha_n}}}\mathrm{d}x,
\end{eqnarray*}
where (a) follows from (\ref{lambda_s}). The first two integrals are bounded so we examine the third integral. Typically, $\alpha_n \leq 10$, therefore
\begin{eqnarray*}
\int\limits_{\mu^{\alpha_n}}^\infty e^{-x^{\frac{2}{\alpha_n}}}\mathrm{d}x \leq \int\limits_{\mu^{\alpha_n}}^\infty e^{-x^{0.2}} = 5 e^{-x^{0.2}}(\mu^{0.8\alpha_n} + 4 \mu^{0.6\alpha_n} + 12 \mu^{0.4\alpha_n} + 24 \mu^{0.2\alpha_n} + 24) < \infty,
\end{eqnarray*}
which satisfies the second conditions and allows us to write $f_{\mathtt{SNR}_{DL,s}}(z)$ as follows
\begin{eqnarray*}
f_{\mathtt{SNR}_{\mathrm{DL},s}}(z)& =& \frac{- \mathrm{d}}{\mathrm{d}z}  \int\limits_0^{\infty} \exp\left(\frac{-z \sigma_s^2 l}{\mathrm{P}_s \psi_s}\right) f_s(l) \mathrm{d}l = -\int\limits_0^{\infty} \frac{\mathrm{d}}{\mathrm{d}z}   \exp\left(\frac{-z \sigma_s^2 l}{\mathrm{P}_s \psi_s}\right) f_s(l) \mathrm{d}l \\
&= &\frac{\sigma_s^2}{\mathrm{P}_s \psi_s} \int\limits_0^\infty l \exp\left(\frac{- z \sigma_s^2 l}{\mathrm{P_s \psi_s}}  \right) f_s(l) \mathrm{d}l.
\end{eqnarray*}
 which concludes this proof and $\mathcal{B}_{UL, m}$ is derived similarly.
\end{proof}
\section{}
\label{Theorem1_proof}
\begin{proof}[\unskip\nopunct]
\textit{Derivation of Theorem 1:}
The DL $\mathtt{SINR}$ coverage for sub-6GHz Mcells is first derived. As shown in Section \ref{Max_RP}, the condition for association to a Mcell in the DL is $L_{\min, s} > a_{DL} L_{\min, m}$.
\begin{eqnarray*}
\begin{aligned}
\mathcal{P}_{DL , m} (\tau)  &=  \mathbb{P} (\mathtt{SINR}_{\mathrm{DL} , m} > \tau ; K_{\mathrm{DL}} = m) = \int\limits_0^{\infty} \mathbb{P}(\mathtt{SINR}_{\mathrm{DL} , m} > \tau; K_{\mathrm{DL}} = m | L_{\min, m} =l) \ f_{m}(l) \mathrm{d}l \\
&=    \int\limits_0^{\infty} \mathbb{P}(\mathtt{SINR}_{\mathrm{DL} , m} > \tau; L_{\min, s} > a_{\mathrm{DL}} l| L_{\min, m} =l) \ f_{m}(l) \mathrm{d}l \\
&\stackrel{(a)}{=} \int\limits_0^{\infty} \mathbb{P}(\mathtt{SINR}_{\mathrm{DL} , m} > \tau| L_{\min, m} =l) \  \mathbb{P}( L_s > a_{\mathrm{DL}} l| L_{\min, m} = l) \ f_{m}(l) \mathrm{d}l \\
&= \int\limits_0^{\infty} \mathbb{P}(\mathtt{SINR}_{\mathrm{DL} , m} > \tau| L_{\min, m} =l) \bar{F}_{s} (a_{\mathrm{DL}} l) f_{m}(l) \mathrm{d}l,
\end{aligned}
\end{eqnarray*}
where (a) follows from the assumption that $\Phi_s$ and $\Phi_m$ are independent.
Since $\bar{F}_{s} $ and $f_{m}$ are known, we now derive the first part of the equation which is given by
\begin{eqnarray*}
\begin{aligned}
\mathbb{P} (\mathtt{SINR}_{\mathrm{DL},m}  > \tau| L_{\min, m} =l)  &= \mathbb{P} \left(\frac{ \mathrm{P}_{m}  \psi_m h_{x^*, 0} l^{-1}}{ I + \sigma_m^2}> \tau | L_{\min, m} = l  \right)\\
& = \exp\left(\frac{-\tau \sigma_m^2 l}{\mathrm{P}_{m}  \psi_m} \right)  \mathbb{E}_I \left[ \exp\left( \frac{-\tau I l}{\mathrm{P}_{m}  \psi_m}  \right) \right]
 =  \exp\left(\frac{-\tau \sigma_m^2 l}{\mathrm{P}_{m}  \psi_m} \right) \mathcal{L}_{I_l} \left(  \frac{-\tau l}{\mathrm{P}_{m}  \psi_m} \right),
\end{aligned}
\end{eqnarray*}
where $\mathcal{L}_I(s)$ is the Laplace transform of the interference and is given by
\begin{eqnarray*}
\begin{aligned}
\mathcal{L}_{I_l}(s)  &= \mathbb{E} [e^{-sI} ] =  \mathbb{E}_{\Phi, h_{x, 0}} \left[ \exp\left(-s\sum\limits_{{{x}} \in {\Phi_{m}}\backslash x^*}\mathrm{P}_{m}  \psi_m h_{x, 0} L_{m}(x)^{-1} \right) \right] \\
&= \mathbb{E}_{\Phi, h_{x, 0}} \left[ \prod\limits_{{{x}} \in {\Phi_{m}}\backslash x^*} \exp\left( -s \mathrm{P}_{m}  \psi_m h_{x, 0} L_{m}(x)^{-1}  \right)\right]\\
&= \exp\left(- \int\limits_l^{\infty} \left(1- \mathbb{E}_{h_{x, 0}} \left[\exp\left( -s \mathrm{P}_{m}  \psi_m h_{x, 0} t^{-1}\right)\right]\right) \Lambda_m(\mathrm{d}t)\right),
\end{aligned}
\end{eqnarray*}
where $\Lambda_m(\mathrm{d}t) $ is given by deriving the expression in (\ref{lambda_m}) with respect to t.
\begin{eqnarray*}
\begin{aligned}
\mathcal{L}_{I_l}(s ) = \exp\left( \frac{-2 \pi \lambda_m}{\alpha_m} \int\limits_l^{\infty} \frac{t^{\frac{2}{\alpha_m}-1}}{1+ \frac{t}{s \mathrm{P}_{m}  \psi_m }} \mathrm{d}t \right).
\end{aligned}
\end{eqnarray*}
Finally,
\begin{eqnarray*}
\begin{aligned}
\mathcal{P}_{\mathrm{DL} , m} (\tau)&= \int \limits_0^{\infty} \exp\left( \frac{-\tau \sigma_m^2 l}{\mathrm{P}_{m}  \psi_m} \right) \exp\left( \frac{-2\pi \lambda_m}{\alpha_m} \int \limits_l^{\infty} \frac{t^{\frac{2}{\alpha_m}-1}}{1 + \frac{t}{\tau l}} dt \right) \bar{F}_s (a_{\mathrm{DL}} l) f_m(l) \mathrm{d}l.
\end{aligned}
\end{eqnarray*}
$\mathcal{P}_{\mathrm{UL} , m}$ is derived in the same way replacing $ \mathrm{P}_{m}$ and $ a_{\mathrm{DL}}$ by $ \mathrm{P}_{um}$ and $ a_{\mathrm{UL}}$ respectively.

We then derive the DL $\mathtt{SNR}$ coverage for mmWave Scells. Similarly, the condition for association to a Scell in the DL is $L_{\min, m} > \frac{L_{\min, s}}{a_{\mathrm{DL}}} $.
\begin{eqnarray*}
\begin{aligned}
\mathcal{P}_{\mathrm{DL} , s} (\tau)  &=  \mathbb{P} (\mathtt{SINR}_{\mathrm{DL} , s} > \tau ; K_{\mathrm{DL}} = s) = \int\limits_0^{\infty} \mathbb{P}(\mathtt{SINR}_{\mathrm{DL} , s} > \tau; K_{\mathrm{DL}} = s | L_{\min, s} =l) \ f_{s}(l) \mathrm{d}l \\
&= \int\limits_0^{\infty} \mathbb{P}(\mathtt{SINR}_{\mathrm{DL} , s} > \tau| L_{\min, s} =l) \bar{F}_{m} \left(\frac{l}{a_{\mathrm{DL}}}\right) f_{s}(l) \mathrm{d}l.
\end{aligned}
\end{eqnarray*}
\begin{eqnarray*}
\begin{aligned}
\mathbb{P} (\mathtt{SNR}_{\mathrm{DL},s}  > \tau| L_{\min, s} =l)  &= \mathbb{P} \left(\frac{ \mathrm{P_{s}}  \psi_s h_{x^*, 0}  l^{-1}}{ \sigma_s^2}> \tau | L_{\min, s} = l  \right)\\
& = \mathbb{P} \left( h_{x^*, 0} > \frac{-\tau \sigma_s^2 l}{\mathrm{P_{s}}  \psi_s  }  | L_{\min, s} = l  \right)
 = \exp\left( \frac{-\tau \sigma_s^2 l}{\mathrm{P_{s}}  \psi_s} \right).
\end{aligned}
\end{eqnarray*}
Finally,
\begin{eqnarray*}
\begin{aligned}
\mathcal{P}_{\mathrm{DL} , s} (\tau)&= \int \limits_0^{\infty}\exp\left( \frac{-\tau \sigma_s^2 l}{\mathrm{P_{s}}  \psi_s} \right) \bar{F}_m \left(\frac{l}{a_{\mathrm{DL}}}\right) f_s(l) \mathrm{d}l.
\end{aligned}
\end{eqnarray*}
$\mathcal{P}_{\mathrm{UL} , s} (\tau)$ is derived similarly by exchanging $\mathrm{P}_s$ and $a_{\mathrm{DL}}$ by $\mathrm{P}_{us}$ and $a_{\mathrm{UL}}$ respectively in the previous derivation.
\end{proof}
\bibliographystyle{IEEEtran}
\bibliography{IEEEabrv,mmw_UHF_DUDe6}
\end{document}